%% file: main.tex
\documentclass[useAMS,usenatbib,fleqn]{mnras}

\usepackage[T1]{fontenc}
\usepackage{ae,aecompl}

\usepackage{times}
\usepackage{subfigure}
\usepackage{amsmath}
\usepackage{amssymb}
\usepackage{amsbsy}
\usepackage{bmpsize}
\usepackage{graphicx}
\usepackage{paralist}
\usepackage{mathrsfs}
\usepackage{booktabs}
\usepackage{tabularx}
\usepackage{courier}
\usepackage{verbatim}
\usepackage{lscape}
\newcommand{\e}{\mathrm{e}}
\newcommand{\percent}{\mathrm{per\>cent}}
\newcommand{\dd}{\mathrm{d}}
\newcommand{\GammaAB}[2]{\Gamma\Big(\frac{#1}{#2}\Big)}
\newcommand{\bs}[1]{\boldsymbol{#1}}
\newcommand{\fJ}{f(\bs{J})}

\title[Models of flattened, rotating globular clusters]{The Gaia-ESO Survey: Dynamical models of flattened, rotating globular clusters}
\author[S.~M.~R.~Jeffreson, J. L. Sanders et al.]{Sarah~M.~R.~Jeffreson,$^{1,2}$\thanks{E-mail: s.jeffreson@uni-heidelberg.de}
Jason L. Sanders,$^1$\thanks{E-mail:jls@ast.cam.ac.uk}
N.~W. Evans,$^1$\thanks{E-mail:nwe@ast.cam.ac.uk}
Angus A. Williams,$^1$
\newauthor
G.~F. Gilmore,$^1$
Amelia Bayo, $^3$
Angela Bragaglia,$^4$
Ettore Flaccomio,$^5$
Richard Jackson,$^6$
\newauthor
Robert D. Jeffries,$^6$
Paula Jofr\'e,$^1$
Carmela Lardo,$^7$
Lorenzo Morbidelli,$^8$
Elena Pancino,$^8$
\newauthor
Simone Zaggia,$^9$\\
$^1$ Institute of Astronomy, Madingley Road, Cambridge, CB3 0HA, United Kingdom\\
$^2$ Astronomisches Rechen-Institut, Zentrum f\"ur Astronomie der Universit\"at Heidelberg, M\"onchhofstrasse 12-14, D-69120 Heidelberg, Germany\\
$^3$ Instituto de F\'isica y Astronomi\'ia, Universidad de Valparai\'iso, Chile\\
$^4$ INAF - Osservatorio Astronomico di Bologna, via Ranzani 1,
40127, Bologna, Italy\\
$^5$ INAF - Osservatorio Astronomico di Palermo G.S.Vaiana,
Piazza del Parlamento 1, I-90134 Palermo, Italy\\
$^6$ Astrophysics Group, Keele University, Keele, Staffordshire ST5 5BG, United Kingdom\\
$^7$ Astrophysics Research Institute, Liverpool John Moores University, IC2, Liverpool Science Park, 146 Brownlow Hill, Liverpool L3 5RF, UK\\
$^8$ INAF - Osservatorio Astrofisico di Arcetri, Largo E. Fermi, 5,
50125, Firenze, Italy\\
$^9$ INAF - Osservatorio Astronomico di Padova, Vicolo dell'Osservatorio 5,
I-35122, Padova, Italy
}
\date{Accepted XXX. Received YYY; in original form ZZZ}

\begin{document}
\label{firstpage}
\pagerange{\pageref{firstpage}--\pageref{lastpage}} \pubyear{2016}
\maketitle
\begin{abstract}
We present a family of self-consistent axisymmetric rotating globular cluster models which are fitted to spectroscopic data for NGC 362, NGC 1851, NGC 2808, NGC 4372, NGC 5927 and NGC 6752 to provide constraints on their physical and kinematic properties, including their rotation signals. They are constructed by flattening Modified Plummer profiles, which have the same asymptotic behaviour as classical Plummer models, but can provide better fits to young clusters due to a slower turnover in the density profile. The models are in dynamical equilibrium as they depend solely on the action variables. We employ a fully Bayesian scheme to investigate the uncertainty in our model parameters (including mass-to-light ratios and inclination angles) and evaluate the Bayesian evidence ratio for rotating to non-rotating models. We find convincing levels of rotation only in NGC 2808. In the other clusters, there is just a hint of rotation (in particular, NGC 4372 and NGC 5927), as the data quality does not allow us to draw strong conclusions. Where rotation is present, we find that it is confined to the central regions, within radii of $R \leq 2 r_h$. As part of this work, we have developed a novel q-Gaussian basis expansion of the line-of-sight velocity distributions, from which general models can be constructed via interpolation on the basis coefficients.
\end{abstract}


\section{Introduction}
Once believed to be among the simplest of astrophysical systems, the complexity of globular clusters is becoming increasingly apparent, as detailed observations reveal clues to their rich evolutionary history. Many globular clusters exhibit a non-negligible degree of flattening~\citep{WhiteShawl1987,ChenChen2010} which appears to vary with radius from the cluster centre~\citep{Geyer1983} and which displays a correlation with internal rotation~\citep{Bellazzini2012,Bianchini2013,Fabricus2014,Kacharov2014}. Multiple stellar populations have been identified, first as chemically distinct features in the spectroscopic light abundance patterns~\citep[e.g.~Na-O anti-correlations][]{Gratton2004,Carretta2010,Gratton2012} and later as distinct photometric features \citep{Piotto2009} in the main sequence~\citep[e.g.][]{Piotto2007,Milone2010,Simioni2016}, the sub-giant branch \citep[e.g.][]{Milone2008} or horizontal branch \citep[e.g.][]{Dalessandro2011}. Recently,~\cite{Cordero2017} has calculated a 98.4\% probability that differential rotation exists between chemically-distinct subpopulations in M13, demonstrating the possibility that multiple populations may also be dynamically-distinguishable.

The use of dynamics in grappling with the origin and evolution of multiple populations is explored in detail in~\cite{Henault-Brunet}. In this work, the rotational signature after a Hubble time is shown to distinguish between the two most prominent scenarios of subpopulation formation, namely the AGB scenario~\citep{Decressin2007,DErcole2008,Conroy2012,Krause2013} and the early disc accretion scenario~\citep{Bastian2013}. Although a large number of discrepancies exist between the predictions of current models and observations~\citep{Henault-Brunet,Bastian2015}, the potential power of combining chemistry and dynamics to explore current and future models of cluster formation and evolution is clear. Additionally, the correlation between rotation and flattening~\citep{Bekki2010,Mastrobuono-BattistiPerets2013,Mastrobuono-BattistiPerets2016}, as seen in high-quality spectroscopic observations of Galactic globular clusters~\citep{Lane2009,Lane2010,Bellazzini2012,Bianchini2013,Fabricus2014,Kacharov2014}, could provide dynamical clues concerning cluster formation and evolution in the Milky Way and other galaxies.

In order to extract the maximum possible information from these increasingly-precise measurements of internal rotation, along with proper motion measurements~\citep{Bellini2014,Watkins2015} and integral field spectroscopy measurements~\citep{Luetzgendorf2012,Bacon2014,Kamann2016}, flexible dynamical models of rotating, flattened globular clusters are required. A number of suitable models have been proposed, including the construction of distribution functions via a modified Schwarzschild orbit superposition method, pioneered by~\cite{vandeVen2006} for $\omega$ Cen and subsequently applied to M15 by~\cite{vandenBosch2006}. This approach allows the degree of solid-body rotation and the inclination angle of the cluster to be simultaneously constrained. Two classes of self-consistent rotating dynamical equilibria are also presented by~\cite{VarriBertin2012}, which control rotation via a set of three parameters within modified~\cite{King1966} models. Further relevant models include non-parametric fits of $\omega$ Cen~\citep{Merritt1997}, truncated Maxwellian models of M13 with solid-body rotation~\citep{Lupton1987}, 2D Fokker-Planck time-evolving models with initial conditions tailored to observations of individual globular clusters~\citep{Fiestas2006}, and the application of~\cite{Wilson1975} models to $\omega$ Cen by~\cite{Sollima2009}.

The construction of distribution functions which depend only upon the canonical set of adiabatically-invariant integrals of motion (action variables) presents an opportunity to increase the flexibility of previous dynamical models \citep{BinneyConfProc}. A spherical isotropic action-based, self-consistent, equilibrium distribution function can be constructed via the method of~\cite{WilliamsEvans2015} and~\cite{Posti2015} to reproduce a density profile suitable for globular clusters. These models can be generalized to axisymmetry (or possibly triaxiality) through the use of approximate action estimation schemes \citep{SandersBinney2016} to produce flattened equilibria. Rotation can be included using the procedure described in~\cite{Binney2014}, whereby internal rotation is independent of the density profile and may be straight-forwardly manipulated subsequent to the computation of density isophotes. As discussed in~\cite{Binney2014}, the linearity of action-based distribution functions allows multiple rotational components to simply be added together without altering the overall density profile, such that these models could feasibly be extended to account for multiple populations with differential rotation. In this paper we present and apply a family of self-consistent dynamical equilibrium models for rotating, flattened globular clusters, noting that these models represent simple building blocks from which multi-component models can trivially be constructed. We argue that the flexibility of these models makes them ideally suited to explore the wide range of phenomena observed in globular clusters today, and ultimately to distinguish between possible scenarios for their formation.

To demonstrate the capability of our action-based dynamical models in constraining the physical properties of globular clusters, we fit the line-of-sight velocity kinematics of six globular clusters from the Gaia-ESO survey, complemented by other archival data. We introduce a novel expansion method to parameterise the model line-of-sight velocity distributions, enabling us to rapidly evaluate the probability of the data for a given set of model parameters. For our six globular clusters, we fit the rotation signal as well as the mass-to-light ratio, inclination angle and systemic velocities and discuss the link between the rotation signal and ellipticity for our sample.

The paper is organised as follows. In Section~\ref{Sec::Data}, we describe the globular cluster data we will use for our modelling. Section~\ref{Sec::Model} outlines how the self-consistent flattened rotating action-based models are constructed and the adopted functional form for our models. In Section~\ref{Sec::DataAnalysis}, we introduce a novel approach to fit a suite of self-consistent models to spectroscopic data. The results of the analysis are discussed in Section~\ref{Sec::Discussion} before we present our conclusions in Section~\ref{Sec::Conclusions}. We also provide two useful appendices. The first gives a number of formulae for the introduced Modified Plummer model and the second details the q-Gaussian basis expansion for line-of-sight velocity profiles, which is employed in the data analysis.

\section{Data}\label{Sec::Data}
We opt to work with spectroscopic and photometric data for the six globular clusters listed in Table~\ref{Table::Summary}. In this section, we describe in turn the sources of the data used and any preliminary processing we have performed.

\begin{table*}
\caption{Summary of properties of globular clusters used in this work. Column (1) gives the total number of spectroscopic targets $N$. Columns (2)-(4) give the number from each of the three sources. Columns (5) and (6) give the 5th and 95th percentiles for the overall radial spread of the data, in units of the half-light radius $r_h$ from~\protect\cite{Harris2010}. Column (7) gives the distance to each cluster $d$. Columns (8) gives the ellipticity measured by ~\protect\citep{WhiteShawl1987} with the fraction of the half-light radius at which the ellipticity was measured in brackets, while Column (9) gives the position angle measurement from the same source. Columns (10) and (11) give the position angle and rotation amplitude, assuming the major axis is perpendicular to the rotation axis determined using the method described in Section~\ref{Section::rotation_case}.
}
\input{Table1.dat}
\label{Table::Summary}
\end{table*}


\subsection{Spectroscopic data} \label{Sec::spectroscopic_data}
Our primary source of spectroscopic data is the Gaia-ESO survey \citep[GES,][]{Gilmore2012,Randich2013}. GES is a public spectroscopic survey using the mid-resolution ($R\sim17000$) FLAMES-GIRAFFE and the higher resolution ($R\sim47000$) FLAMES-UVES spectrographs on the VLT. For most clusters, the bulk of the spectra use the combination of the HR10 (534-562 nm) and HR21 (848-900 nm) setups on GIRAFFE although $\sim60\,\percent$ of the GIRAFFE data for NGC 6752 and  $\sim40\,\percent$ for NGC 362 use the HR15N setup (647-679 nm). There are a small number of spectra using the UVES U580 (480-680 nm) setup (and 3 for NGC 6752 using the U520 setup, 420-620 nm): 6 for NGC 362, 26 for NGC 1851, 19 for NGC 2808, 1 for NGC 4372, 6 for NGC 5927 and 21 for NGC 6752. We use the metallicities and radial velocities given in the fourth internal data release (iDR4). We have combined this data with data from~\cite{Lardo2015}\footnote{These authors use GES data combined with archival ESO data, so when we use the term `Lardo data' we mean all data used by \cite{Lardo2015} but not in GES.},~\cite{Lane2011} (for NGC 6752) and~\cite{Kacharov2014} (for NGC 4372). The spectroscopic sample was formed by combining the stars from each of these sources and removing duplicates (defined as a star within an on-sky distance of two arcseconds from another star in our sample -- we preferentially retain the star with the lower radial velocity uncertainty). $3\sigma$ outliers in metallicity were removed from the Gaia-ESO sample, as metallicity was unavailable for the other datasets (but the authors do provide flags for likely membership which we use instead). Additionally, we remove all stars outside $12$ scale radii of our models (defined later) as these fall outside our model grids, and all stars outside $v_r\pm4\sigma_v$, where $v_r$ and $\sigma_v$ are the mean velocity and velocity dispersion reported in~\citet[2010 edition]{Harris2010}. For NGC 4372 and NGC 5927, we use a default value of $\sigma_v=5\,\mathrm{km\,s}^{-1}$ as no value is provided by~\citet[2010 edition]{Harris2010}. The presence of binaries (or other multiple systems) may affect our results, but only in NGC 6752 has a single spectroscopic binary been detected from the Gaia-ESO data (Merle et al., in prep.) so binarity is not important for our sample. Furthermore, \cite{Milone2012} has measured the binary fraction in NGC 362, NGC 1851, NGC 5927 and NGC 6752 using photometry and found that the binary fraction is approximately less than $5\percent$ for these clusters (although for a core sample in NGC 5927, the binary fraction was found to be $10\percent$).

\begin{figure}
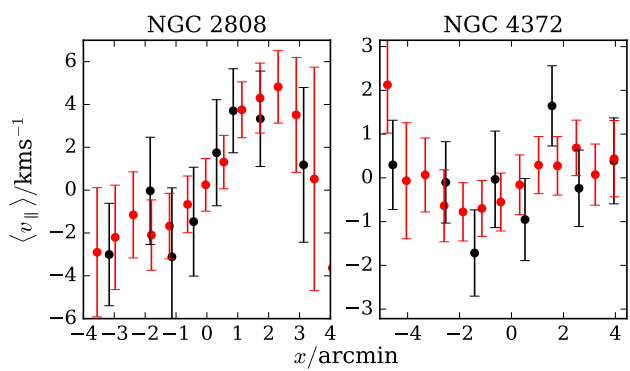

$$\includegraphics[width=\linewidth]{{{figs/Fig1}}}$$
\caption{Rotation curves for NGC 2808 and NGC 4372 obtained from the spectroscopic data using the procedure described in Section~\ref{Section::rotation_case}. The red points show the mean velocity in overlapping bins along the axis on the sky that maximises the rotation signal, whilst the black points show the result for non-overlapping bins.}
\label{Fig::rotation_curves}
\end{figure}

We transform our entire sample to on-sky coordinates aligned with the major axis of the cluster $(x,y)$ by first converting the on-sky angular positions to a local Cartesian basis \citep[using equation (1) from][]{vandeVen2006} and then rotating by the position angle. We primarily use the position angle measurements of the major axis from \cite{WhiteShawl1987} (see the following sub-section), except for NGC 4372, where we use the value reported by~\cite{Kacharov2014}, and NGC 362, where we choose the position angle according to the rotation axis which maximises the rotation signal, via the procedure implemented by the same authors (and described in the following sub-section). We opt to use this technique for NGC 362 due to its near-sphericity and the consequent ambiguity of its position angle. This ambiguity is reflected in the large disparity between the values reported in~\cite{WhiteShawl1987} and~\cite{ChenChen2010}, which are $25 \pm 4$ and $61 \pm 7\deg$, respectively.

For visualization purposes only, the mean velocities and total velocity dispersions are fitted using a Gaussian mixture model (GMM) to prevent skewing of the distribution by any further outliers. 

\subsubsection{The case for rotation}\label{Section::rotation_case}

Several of the clusters in our sample display evidence for internal rotation. To obtain a preliminary indication of whether rotation is present, we follow the procedure outlined in \cite{Bellazzini2012} and~\cite{Kacharov2014}. The rotation axis is determined by measuring the difference in mean line-of-sight velocities, $\Delta\langle v_{||}\rangle$, on either side of a given axis orientated at $18$ position angles, $\phi$, separated by $20\,\mathrm{deg}$ and their corresponding Poisson errors. The resulting run of $\Delta\langle v_{||}\rangle$ with $\phi$ is fitted with a sinusoid $-A_\mathrm{0}\cos(\phi-\mathrm{PA}_0)$ to find the position angle of the rotation axis, $\mathrm{PA}_0$, for which the rotation signal is maximised\footnote{The sinusoidal amplitude $A_0$ should not be confused with $A_\mathrm{rot}$, which is simply the maximum amplitude of the projected mean velocity profile, independent of the procedure described here. In Section~\protect\ref{Sec::Rotation} we compute the value of $A_\mathrm{rot}$ predicted by our models.}. For an axisymmetric model, the major axis is orthogonal to the rotation axis, and we then denote the position angle of the major axis as $\mathrm{P.A.}_\mathrm{rot}=\mathrm{PA}_0-\pi/2$. We use the MCMC package \emph{emcee} \citep{emcee} to sample the posteriors on the parameters $A_0$ and $\mathrm{P.A.}_\mathrm{rot}$ (assuming, incorrectly, that the uncertainties in $\Delta\langle v_{||}\rangle$ are uncorrelated) and give the results in Table~\ref{Table::Summary}. In Fig.~\ref{Fig::rotation_curves}, we show the rotation curves for NGC 2808 and NGC 4372, the two most convincing cases. The data points have again been computed using a GMM, with bins along the axis on the sky perpendicular to the rotation axis. We have used both overlapping (red) and non-overlapping (black) bins. The overlapping bins are of width $2\arcmin$ and only include the innermost $90\,\percent$ of the data in $b$. The non-overlapping bins are equally-populated, and we have used four bins on either side of the axis, after removal of the outermost $5\, \percent$ of data on each side. Later in the paper, we will identify rotation by fitting full dynamical models to the data.

\subsection{Ellipticity measurements}
\label{Sec::Ellipticity}
There are two studies of the flattening of globular clusters from~\citet{WhiteShawl1987} and~\cite{ChenChen2010}. While~\cite{ChenChen2010} use infrared data, so are less susceptible to extinction, they consider the flattening at significantly larger radii than~\cite{WhiteShawl1987} in all cases, where Poisson noise and tidal effects begin to dominate. This is emphasised by the disagreement in the position angles reported by the two sets of authors. We opt to work with the data from~\citet{WhiteShawl1987} in all cases but two. For NGC 4372, we use the ellipticity $\epsilon = 0.08$ and position angle $\mathrm{P.A.}=48\,\mathrm{deg}$ derived by~\cite{Kacharov2014}. For NGC 362, we choose to use the position angle that maximises the rotation signal $(\mathrm{P.A.}_\mathrm{rot}=131\deg)$ and use the \cite{WhiteShawl1987} value for the ellipticity. As we do not have full ellipticity profiles for each globular cluster, and since we primarily use the data from \citet{WhiteShawl1987}, we use our models to reproduce only the flattening near the scale radius, rather than fitting the ellipticity at all radii.


\subsection{Photometric data}

\begin{figure*}
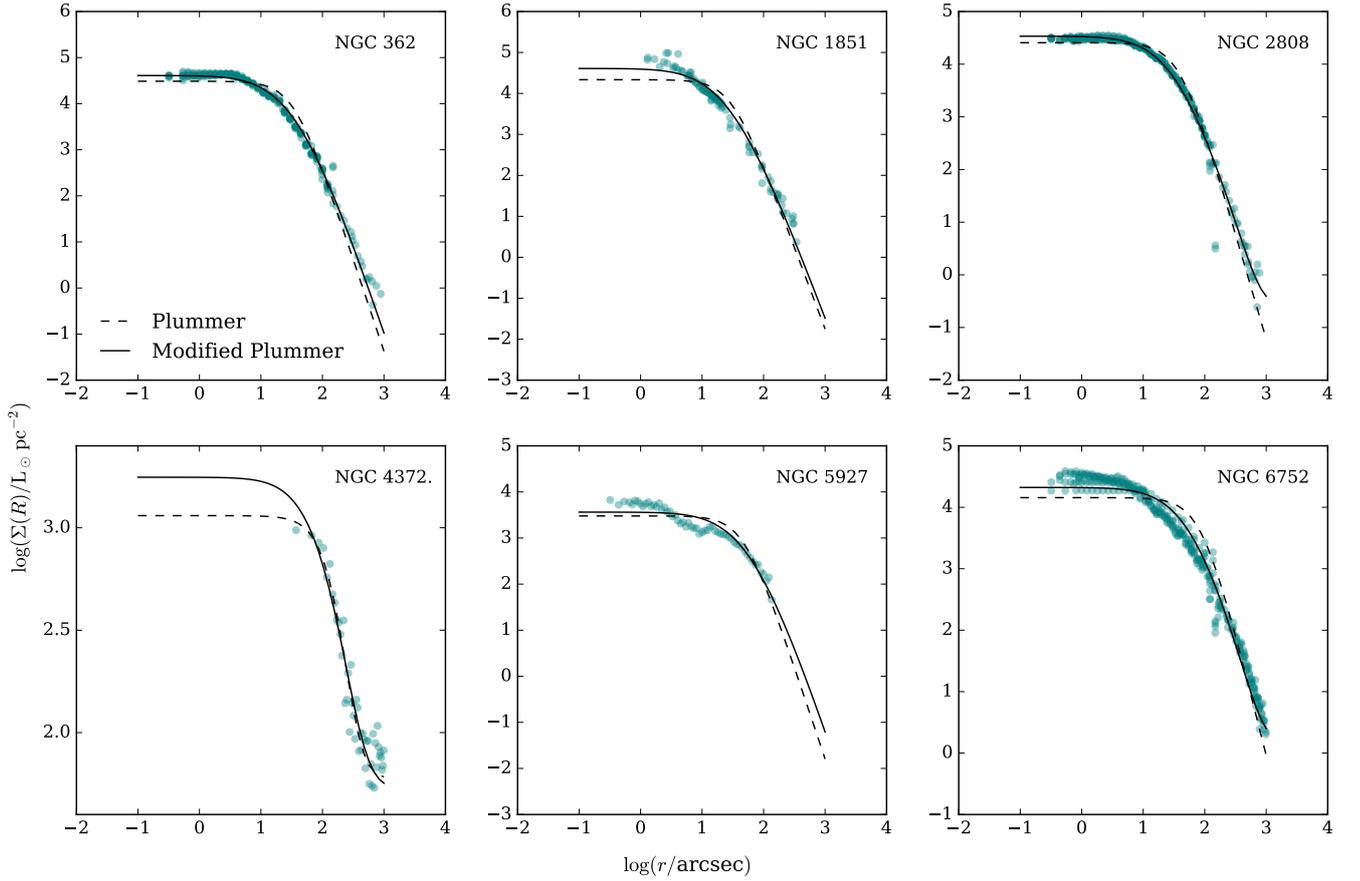

$$\includegraphics[width=\textwidth]{{{figs/Fig2}}}$$
\caption{Fits of the projected density for our models to the photometric data from~\protect\cite{Kacharov2014} for NGC 4372 and the surface brightness profiles from~\protect\cite{Trager1995} for the other five clusters. Note the improved fits of the transition region given by the Modified Plummer model (solid black lines) relative to the Plummer model (dashed black lines) for the clusters NGC 362, NGC 1851, NGC 2808 and NGC 6752.}
\label{Fig::DensityProf}
\end{figure*}

\begin{table}
\caption{Surface density fitting results: the results of fitting the projected density profile of the Modified Plummer Model~\eqref{Eqn::NewPlummer} to the surface density profiles of the globular clusters. Column (1) gives the fitted luminosity in solar luminosities, while column (2) gives the central density in solar luminosities per cubic parsec, calculated using the mass within a sphere of radius $0.5 r_s$. Column (3) gives the fitted scale radius in arcmin, while column (4) gives the physical 2D projected half-light radius in units of parsec (related to $r_s$ by $R_h=1.2038r_s$).}
\begin{tabular}{@{}l c c c c @{}}
  \hline
   NGC & $L$ & $\rho_0$ & $r_\mathrm{s}$ & $R_h$ \\
   &[$10^5 L_{\odot}$]&$[\log_{10}(L_\odot/\,\mathrm{pc}^3)]$&$[\mathrm{arcmin}]$&$[\mathrm{pc}]$ \\
   & (1) & (2) & (3) & (4) \\
  \hline
   362  & 1.57 & 4.75 & 0.55 & 1.66 \\
   1851 & 1.68 & 4.73 & 0.41 & 1.74 \\ 
   2808 & 2.04 & 4.57 & 0.62 & 2.08 \\ 
   4372 & 1.19 & 2.74 & 3.40 & 6.91 \\ 
   5927 & 0.30 & 3.53 & 0.89 & 2.40 \\ 
   6752 & 0.83 & 4.45 & 1.17 & 1.64 \\ 
  \hline
  \end{tabular}
\label{Table::FitResults}
\end{table}

We determine the mass and length scales of the globular clusters NGC 362, NGC 1851, NGC 2808, NGC 5927 and NGC 6752 by fitting spherical density profiles to the surface density profiles from \citet{Trager1995}. For NGC 4372, these data are poorly-constrained and so we use instead the deeper resolved data provided by \citet{Kacharov2014}. A surface brightness profile for NGC 4372 is constructed by binning the \citet{Kacharov2014} data in $50$ equally populated circular bins. For each surface brightness profile, we fit the surface density profile
\begin{equation}
\Sigma(R) = \Sigma_\mathrm{model}(R)+n_f,
\end{equation}
where $\Sigma_\mathrm{model}(R)$ is the model surface density profile and $n_f$ is the contribution from background stars (only important for NGC 4372).
Although Plummer profiles provide reasonable fits to the surface density profiles of old globular clusters, they are not as well suited to the clusters in our sample, four of which are classified as `young' clusters, based on the positions of their main-sequence turn-offs and age-metallicity relations~\citep{MarinFranch2009}. 

Figure~\ref{Fig::DensityProf} shows the photometric data, together with Plummer fits as dashed  lines.
A significantly better fit for the clusters NGC 362, NGC 1851, NGC 2808 and NGC 6752 is provided by a profile with a Plummer-like behaviour at large and small radii, but with a slower turnover.  A simple model that possesses this property has mass density
\begin{equation}
\rho(r)=\frac{3M}{\pi r_s^3}\Big(1+\frac{r}{r_s}\Big)^{-5}.
\label{Eqn::NewPlummer}
\end{equation}
We will call this the Modified Plummer Model. In Appendix~\ref{Appendix::SurfaceDensity}, we provide the analytic expression for the surface density of this model. This model has a 3D half-light radius of $r_h=1.5925r_s$ and a 2D projected half-light radius of $R_h=1.2038r_s$. The surface brightness profile of this model is shown as solid lines in  Figure~\ref{Fig::DensityProf}.  We  provide the computed normalization and scale radius using a least-squares approach, (not accounting for the uncertainties as we assume we are systematic dominated) in Table~\ref{Table::FitResults}. The remaining two `older' clusters NGC 4372 and NGC 5927~\citep{Kacharov2014,MarinFranch2009}, are equally well fitted by either the modified or  traditional Plummer profile, thus we opt to use the former to model all clusters in the sample. 

\section{Modelling Framework}\label{Sec::Model}

\begin{figure*}
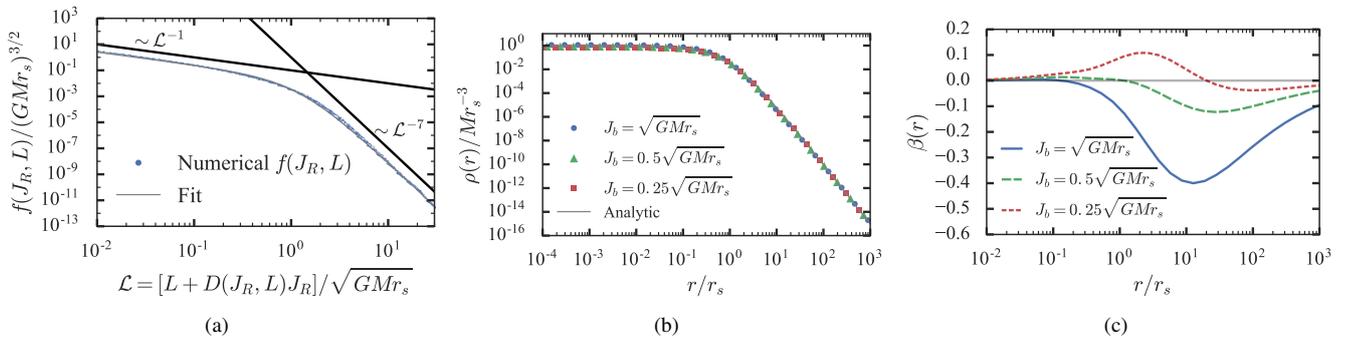

\subfigure[]{\label{A}\includegraphics[width=0.33\textwidth]{{{figs/Fig3a}}}}
\subfigure[]{\label{B}\includegraphics[width=0.33\textwidth]{{{figs/Fig3b}}}}
\subfigure[]{\label{C}\includegraphics[width=0.33\textwidth]{{{figs/Fig3c}}}}
\caption{The numerically-constructed action-based distribution function. Panel (a) shows the numerically-constructed distribution function at a series of $J_R$ and $L$ as blue dots with the grey line showing the analytic fit and the black lines showing the asymptotic limits. The middle panel (b) shows the density profile for this model with three values of the anisotropy scale $J_b$, corresponding to different anisotropy profiles shown in the right panel (c), where $\beta = 1-\sigma^2_{tt}/2\sigma^2_{rr}$.}
\label{fJdots}
\end{figure*}

In this section, we lay out the modelling framework used to analyse the globular cluster data. For stars in the cores of globular clusters, the relaxation time can be of order the Hubble time (i.e. the age of the cluster), such that the evolution of globular clusters is shaped by collisional processes. However, the present state should be well modelled by collisionless dynamics, as the orbital time-scale is much shorter than the relaxation time. We assume that the globular clusters are optically thin with no dark matter component, and for simplicity we restrict the models to axisymmetry.

Collisionless dynamical equilibria follow a distribution function (DF) $f(\bs{x},\bs{v})$ that must obey the collisionless Boltzmann equation
\begin{equation}
\frac{\dd f}{\dd t}=0.
\end{equation}
By Jeans' theorem, the collisionless Boltzmann equation is satisfied if the DF is solely a function of integrals of motion. One reasonable choice for the integrals of motion are the action variables $\bs{J}$. Along with the angle variables $\btheta$, the action variables form a canonical set of coordinates. Therefore, they represent a very simple way of describing the orbital motion. Additionally, their use is advantageous for our application as the actions are adiabatic invariants and form a Cartesian basis for describing the space of orbits i.e. the range of possible values of one of the actions is independent of the other two \citep[see][for more discussion of the merits of working with action variables]{BinneyTremaine}.

In axisymmetric potentials, the action variables are usually written as $(J_R,J_\phi,J_z)$ where $J_\phi=L_z$ is the component of angular momentum about the symmetry axis, $J_R$ describes the extent of radial motion and $J_z$ the extent of vertical motion. These actions are the natural extension of the actions $(J_r,L_z,L-|L_z|)$ in a spherical potential for an orbit with angular momentum $L$. In general axisymmetric potentials, numerical integrations of orbits produce regular surfaces of section for large regions of phase-space indicating that the action variables exist. However, the most general potential in which the Hamilton-Jacobi equations are separable and thus integrable are the St\"ackel potentials. Numerous algorithms for estimating the actions in general potentials have been proposed \citep[see][for a summary]{SandersBinney2016} with the most rapid being based on the equations of motion in a St\"ackel potential \citep[e.g.][]{Sanders2012,Binney2012,SandersBinney2015}. In brief, these algorithms construct a St\"ackel potential that closely represents the true potential and the actions are estimated as those in the St\"ackel potential. This approach also offers the possibility of significant speed-ups, as we also have access to an approximate third integral $I_3$ which can be used to find the actions via interpolation \citep[see the appendix of][]{Binney2014}. Throughout this paper, we use the axisymmetric action estimation scheme from \citet{Binney2012} with the $\Delta$ estimation scheme from \citet{Binney2014}.

With the advances in action estimation there have also been advances in the construction of appropriate action-based DFs. \cite{WilliamsEvans2015} and \cite{Posti2015} have recently proposed similar spherical action-based DFs for dynamically-hot galactic components such as the Galactic halo \citep{WilliamsEvans2015b,DasBinney2016}, the Galactic dark-matter halo \citep{Piffl2015} or the stellar distribution in giant ellipticals \citep{Posti2016}. Beyond the spherical regime, \cite{Binney2014} has demonstrated how the isotropic isochrone DF can be deformed to produce flattened isochrone models with tangential or radial anisotropy by re-weighting the DF on surfaces of constant energy. In a similar vein, \cite{SandersEvans2015} have constructed triaxial models by rescaling the actions in a spherical DF.

With an action-based DF, the self-consistent solution must be constructed through an iterative procedure. An initial guess of the potential $\Phi_0(\bs{x})$ is made and the density of the model is computed on a grid in this potential. For an action-based DF, the density is simply computed as
\begin{equation}
\rho(\bs{x})=\int\dd\bs{v}\,f(\bs{x},\bs{v})=\int\dd\bs{v}\,f(\bs{J}(\bs{x},\bs{v})).
\end{equation}
Poisson's equation is solved for the corresponding potential $\Phi_1(\bs{x})$ via a multipole expansion \citep[see][for details on the exact implementation]{SandersEvans2015} and the density is computed in this new potential. This process is repeated until the difference between consecutive potential iterations is smaller than some threshold.

\subsection{Construction of suitable DF}

Given a density profile, the procedure outlined by \cite{WilliamsEvans2015} can now be used to construct an appropriate action-based DF $\fJ$ that is approximately isotropic and recovers the required density profile. As touched upon by \citeauthor{WilliamsEvans2015} (\S4.1.2), the procedure appears to break down for cored models with intermediate outer slopes. In general, cored density profiles receive significant contributions in the core from orbits that have actions greater than the scale action, in contrast to cuspy density profiles where larger action orbits only weakly contribute to the central cusp. Thus for cored models with intermediate outer slopes, the assumption that the central regions can be treated independently to the outer regions is no longer valid. The algorithm needs refinement for cored profiles. Here we adopt a purely numerical approach to constructing the DF \citep{BinneyConfProc}.

We begin by using Eddington inversion to construct the $f(E)$ model corresponding to the density profile $\rho(r)$ of the Modified Plummer Model,
\begin{equation}
f(E) \propto \frac{\mathrm{d}}{\mathrm{d}E}\int_0^{-E}\frac{\mathrm{d}\Phi}{\sqrt{\Phi-E}}\frac{\mathrm{d}\rho}{\mathrm{d}\Phi}.
\label{Eqn::EddInv}
\end{equation}
Given a density profile, we compute the potential, $\Phi(r)$, on a logarithmically-spaced grid between $r_\mathrm{min}=10^{-3}r_s$ to $r_\mathrm{max}=10^3 r_s$ using equation (2.28) from \cite{BinneyTremaine} (extrapolation for $r<r_\mathrm{min}$ is performed using a quadratic fit to the three innermost grid points and extrapolation for $r>r_\mathrm{max}$ is performed using a Keplerian fall-off assuming all the mass is contained within $r_\mathrm{max}$.). We perform the differentiation under the integral sign of equation~\eqref{Eqn::EddInv} as given in equation (4.46b) of \cite{BinneyTremaine}. $f(E)$ is computed on a linearly-spaced grid in energy between $\Phi(r_\mathrm{min})$ and $\Phi(r_\mathrm{max})$. The density and potential are interpolated on a logarithmically-spaced grid in radius (between $\tfrac{1}{2}r_\mathrm{min}$ and $2r_\mathrm{max}$) using cubic splines \citep[][for which the first and second derivatives are simply computed]{GSL}. Derivatives of the density with respect to the potential are transformed to derivatives with respect to radius and the integrals are computed using an adaptive Gauss-Kronrod algorithm \citep{GSL} via a change of variables $x=2\sqrt{\Phi-E}$.

We then numerically construct the Hamiltonian as a function of the actions $(J_R,L)$ by interpolating on a grid of $J_R(E,L)$ and $L$. For a logarithmically-spaced grid in the radius $r$ between $r_\mathrm{min}$ and $r_\mathrm{max}$, we compute the angular momentum $L$ and energy $E_c$ of a circular orbit at each radial grid-point. At each radial grid-point, we construct a grid in energy as $E=E_c(1-I)^2$ where $I$ is a linearly-spaced grid in the interval zero to one. For each $r$ and $E$, we store the radial action as
\begin{equation}
J_R = \frac{1}{\pi}\int^{r_a}_{r_p} \mathrm{d}r\,\sqrt{2E-2\Phi(r)-\frac{L^2}{r^2}},
\end{equation}
where $r=r_p,r_a$ are the roots of the integrand and the integral is computed using Gauss-Legendre quadrature \citep{GSL}. Now, given an angular momentum $L$ and radial action $J_R$, we compute the Hamiltonian $H(J_R,L)$ by linear interpolation in $\log J_R$ and $\log L$. Combining this function with $f(E)$ gives us the exact isotropic $\fJ$. This corresponds to a value $\beta = 0$ of the anisotropy parameter

\begin{equation}
\beta = 1-\frac{\sigma^2_{tt}}{2\sigma^2_{rr}},
\end{equation}
where $\sigma_{rr}$ and $\sigma_{tt}$ are the radial and tangential velocity dispersions respectively.

\begin{figure*}
$$\includegraphics[width=\linewidth]{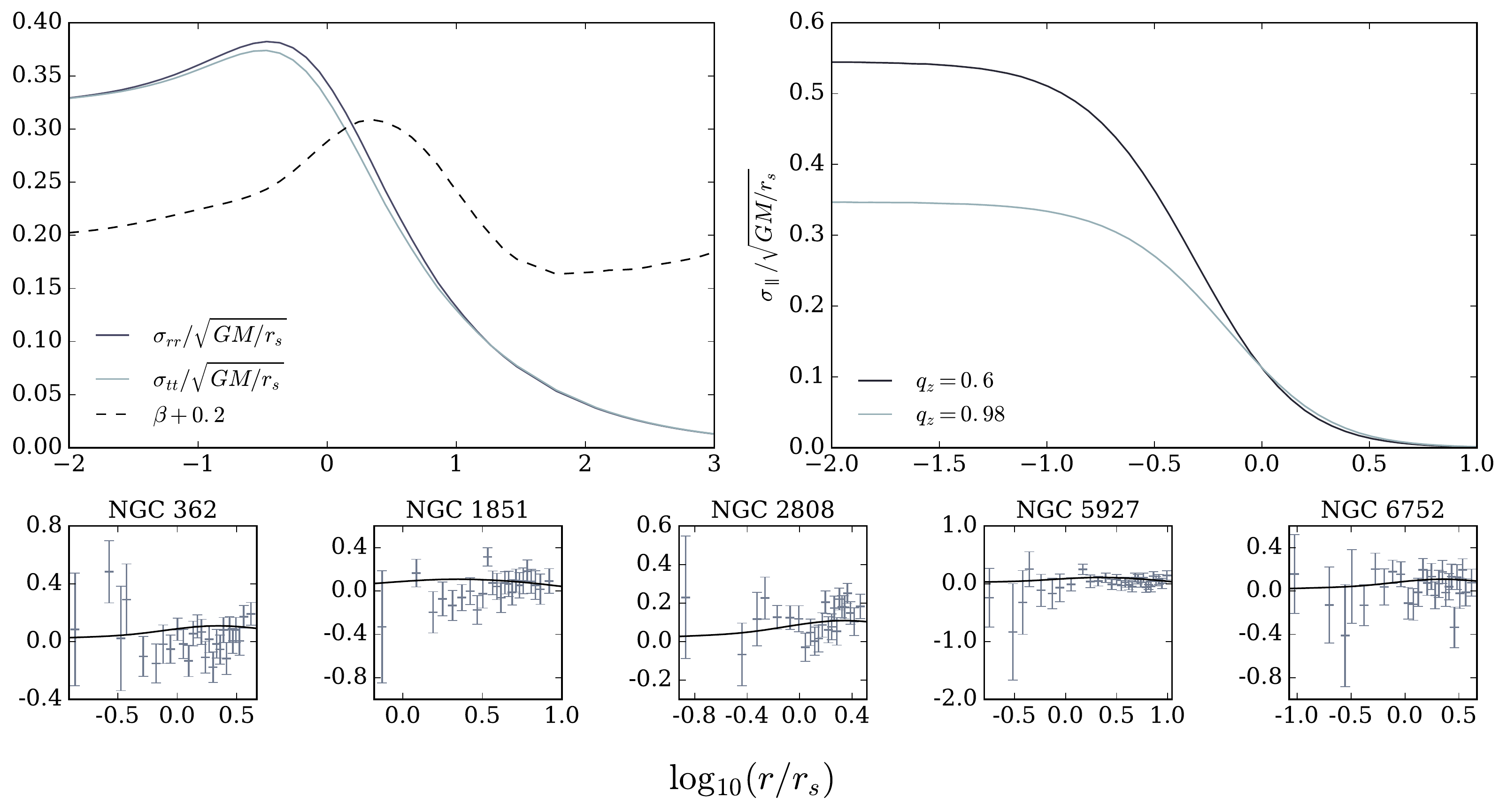}$$
\caption{Left: Intrinsic radial and tangential velocity dispersions produced by the spherical model, along with the value of the anisotropy parameter $\beta=1-\sigma_{tt}^2/2\sigma_{rr}^2$. Note that the anisotropy has a maximum magnitude of around $0.1$. Right: Projected velocity dispersions produced by the models of most extreme flattening parameter ($q_z=0.6$) and least extreme flattening parameter ($q_z=0.98$) used in this work. Bottom: Comparison of the intrinsic anisotropy to the observed values of anisotropy measured in~\protect\cite{Watkins2015}, where these values are available. Note that our anisotropy provides a good fit to these values, in particular it is nowhere larger than the observed anisotropy, such that our spherical model is near-isotropic.}
\label{Fig::Anisotropy}
\end{figure*}

Naturally, it is awkward to work with this purely numerically constructed $\fJ$ so we seek a more compact representation by finding a fitting function that well describes the model. We know that the Hamiltonian tends to that of a harmonic oscillator in the centre and the Keplerian limit at large radii. Therefore, the isotropic DF is a function of $(2J_R+L)$ in the centre and $(J_R+L)$ at large radii. Similar to \cite{WilliamsEvans2015} we suppose that our DF is a function of the variable $\mathcal{L}=D(\bs{J})J_R+L$ where
\begin{equation}
D(\bs{J})=\frac{2+\bs{J}/J_b}{1+\bs{J}/J_b}.
\label{Eqn::DJ}
\end{equation}
This approximation is obviously poorest around the action $J_b$ and a prudent choice of $J_b$ will reduce the degree of anisotropy.

In Figure~\ref{fJdots}, we show the distribution of $f$ as a function of $\mathcal{L}$. We see that it is well represented by a double power-law with an inner slope of $\lambda=1$ and outer slope of $\mu=7$. Therefore, we propose the fitting function
\begin{equation}
\fJ \propto \mathcal{L}^{-1}(J_0^\nu+\mathcal{L}^\nu)^{-(\mu-\lambda)/\nu},
\end{equation}
where we have introduced the additional two variables $J_0$ and $\nu$. By minimising the sum of the square of the differences in the logarithm of $f$ and our model, we find that $J_0\approx1.2\sqrt{GMr_s}$ and $\nu\approx1.6$. For this minimization we set $J_b=J_0$ for which the model is tangentially-biased (see right panel of Fig.~\ref{fJdots}). We therefore have a simple action-based DF that reproduces our required density profile (see the comparison in the central panel of Fig.~\ref{fJdots}).

\cite{Watkins2015} provides anisotropy profiles from Hubble Space Telescope proper motion data for five of the globular clusters studied here. These authors find generically that globular clusters have near isotropic cores and are weakly radially anisotropic near the half-mass radius. We therefore opt to work with near isotropic models. Figure~\ref{fJdots} shows that setting $J_b/J_0=\tfrac{1}{4}$ produces an equally good fit to our numerical $\fJ$ whilst producing a nearly isotropic model with small (radially-biased) fluctuations in the anisotropy near the scale radius. In Fig.~\ref{Fig::Anisotropy} we show the anisotropy profiles from \cite{Watkins2015} for the five globular clusters in common with our study, along with the anisotropy of our fiducial spherical model. The match is very satisfactory. It would be simple to extend our approach to account for differing anisotropy profiles. However, we have found that more radially anisotropic models can develop undesirable prolate cores which we will discuss later.

\subsection{Flattening}

\begin{figure}
$$\includegraphics[width=\columnwidth]{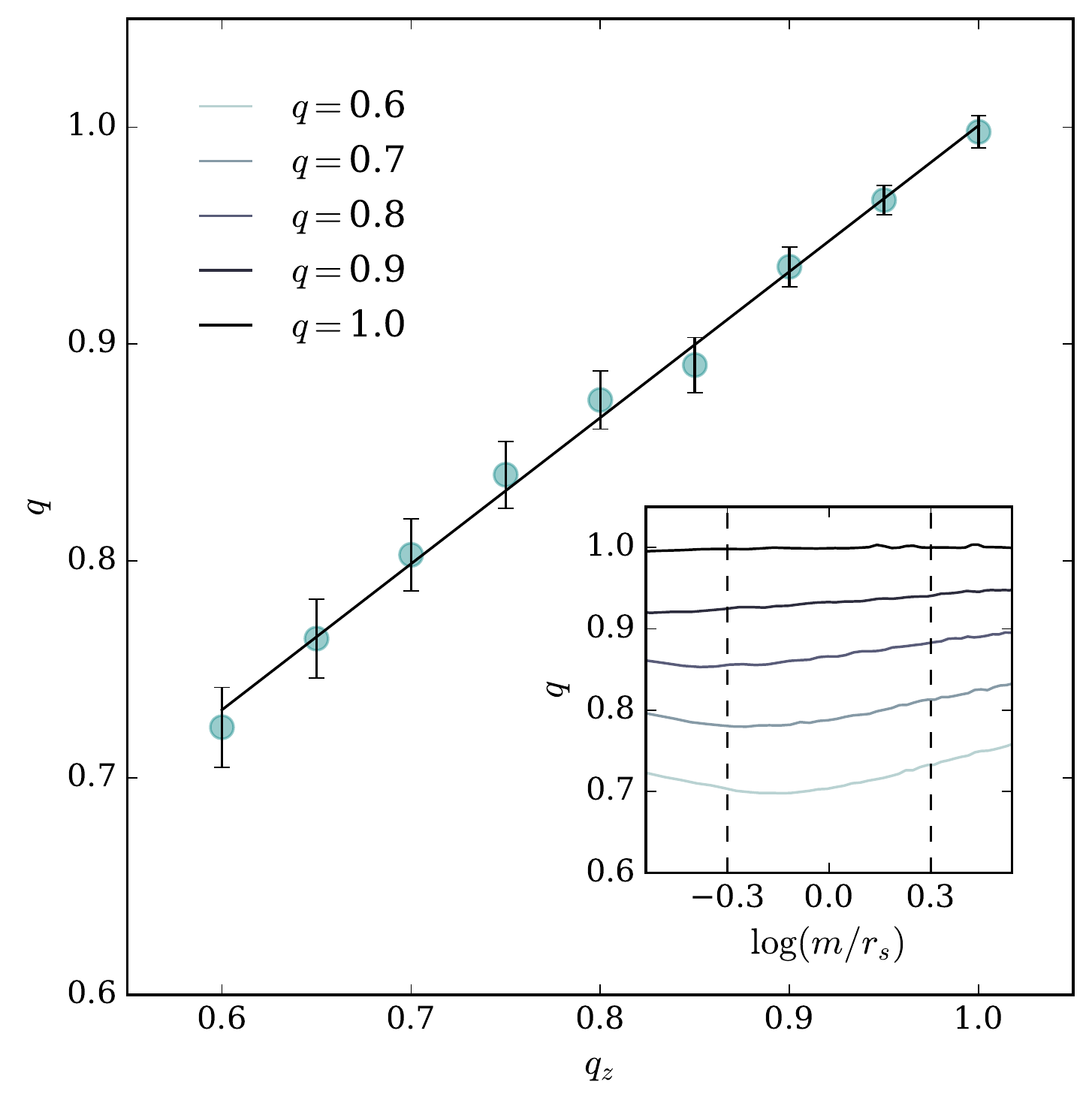}$$
\caption{Linear fit to the relationship between parameter $q_z$ controlling the scale of the longitudinal action in our models, and actual flattening of the model $q$, estimated by fitting 100 ellipses on a logarithmic scale between the radii of $R= \tfrac{1}{2}r_s$ and $R=3r_s$. The error bars represent the standard deviation in the axial ratios of the fitted ellipses. Inset: The change in $q$ with average distance from the cluster centre $m$, for five values of $q_z$. Note that the ellipticity is maximal about the scale radius and the models gradually become more spherical at larger radii.}
\label{Fig::q_against_qz}
\end{figure}

The presented model is spherical as the DF depends only on the angular momentum and the radial action. For our purposes we require the models to be weakly flattened so we must make the DF a function of the vertical action $J_z$. We do this by making the replacement
\begin{equation}
L\rightarrow |J_\phi| + \frac{J_z}{q_z} \text{ where } q_z<1.
\end{equation}
This naturally reduces the weight of high $J_z$ orbits, causing the density profile to flatten whilst retaining the approximate radial profile of the model. $q_z=1$ recovers the spherical DF. In Fig.~\ref{Fig::q_against_qz}, we plot the axial ratio $q$ of ellipses fitted to the density isophotes of our models close to the scale radius $r_s$ against the value of the flattening parameter $q_z$, in the $(R,z)$ plane. As anticipated, there is a linear relationship between the two quantities although the gradient is not unity but better approximated by $q\approx\tfrac{1}{2}(1+q_z)$. This gives us a simple way of relating our model parameter $q_z$ to the observable quantity $q$.

\begin{figure}
\mbox{$$\includegraphics[width=\columnwidth]{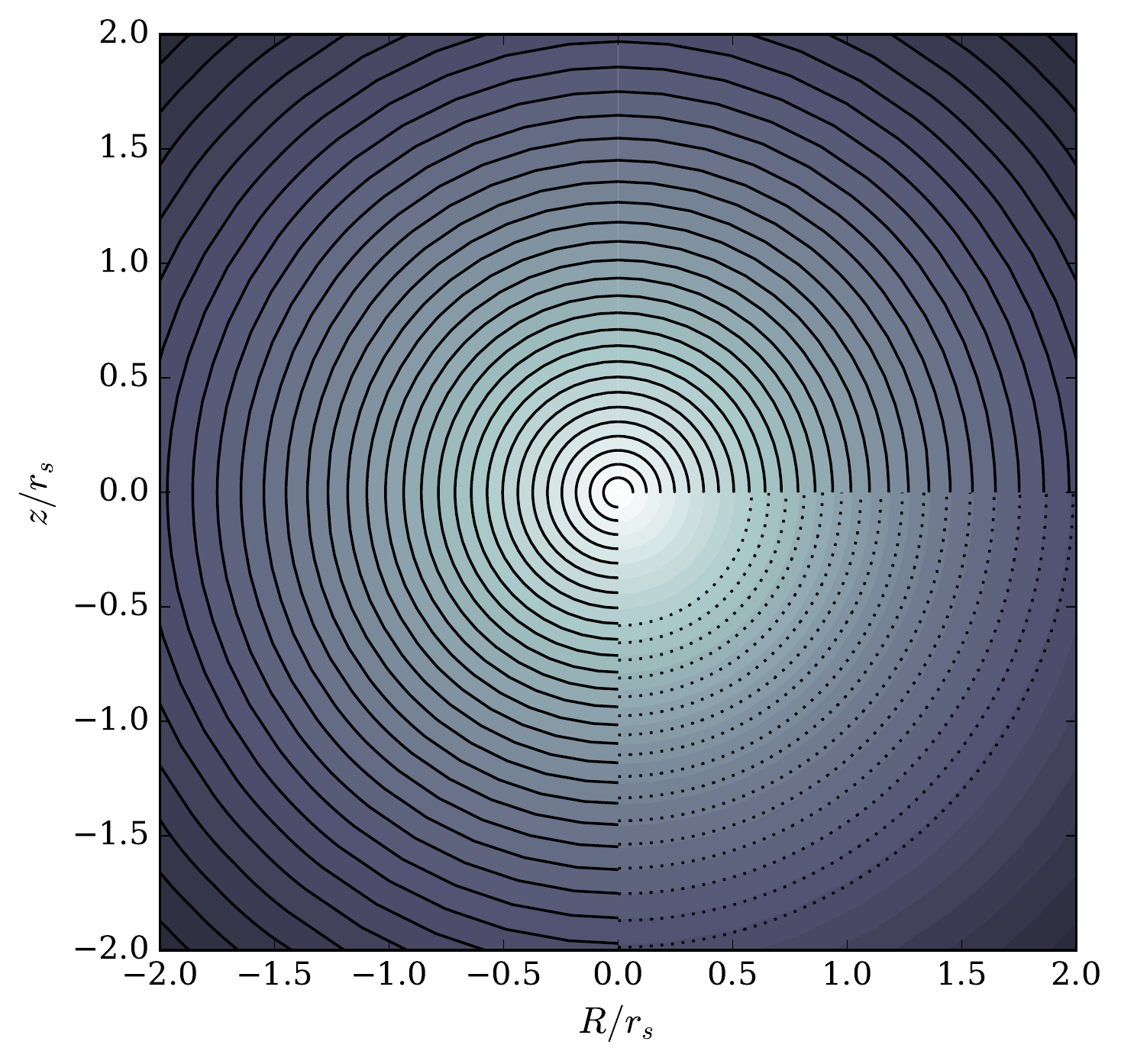}$$}
\mbox{$$\includegraphics[width=\columnwidth]{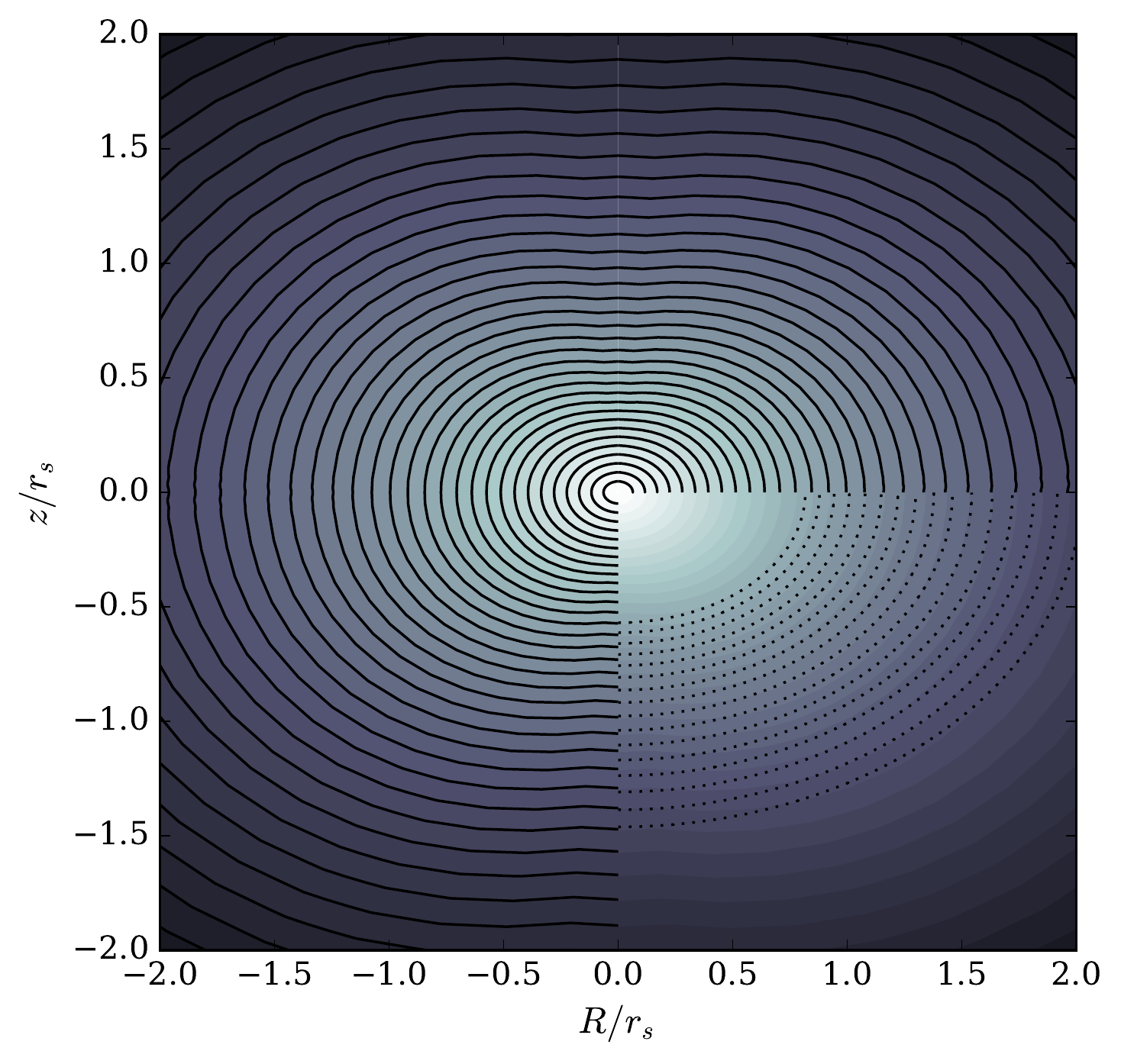}$$}
\caption{Logarithmically-spaced density contours for the spherical case $q_z=1$ (top) and the most flattened case we have used in this work, $q_z=0.6$ (bottom). Solid black lines represent the density contours while dotted black lines show a sample of ellipses fitted on a log scale about the scale radius. Each plot is an axisymmetric plane where the z-axis is vertical.}
\label{Fig::2Dmodelproperties}
\end{figure}

In Fig.~\ref{Fig::2Dmodelproperties}, we show the density contours of the model for the spherical $q_z=1$ and most flattened $q_z=0.6$ cases used in this work. We calculate the density contours on a logarithmic grid in the spherical polar co-ordinate $r$ and over 13 angles $\theta$ from the $z$-axis, ranging from $\theta=0$ to $\theta=\pi/2-0.1$. We do not calculate density values at the north or south poles due to the singularity in the radial action at these points. In our models, we generically see a weak pinching in the density contours near the poles.

As an aside, we note that our simplistic procedure for flattening the specific Modified Plummer model seems to produce physically reasonable models. However, we have found that more radially-biased initial spherical models can produce prolate cores when flattened in the way described. A similar effect was also found by \cite{Binney2014} when studying flattened isochrone models. There appears to be a subtle interplay between the vertical and radial actions that produce the flattening of the model. Near the meridional plane, flattening is simply produced by down-weighting high vertical action orbits. However, near the symmetry axis the situation is more complicated as the density structure is governed primarily by the near-shell orbits. The vertical flattening near the axis is determined by the typical ratio of the radial to vertical action of the contributing orbits. Therefore, models that are more radially-biased in the plane can produce more prolate density contours near the symmetry axis. Whilst our modelling procedure is sufficient for our purposes, it may require more careful thought to extend it to a broader range of density and kinematic profiles.

\subsection{Rotation profiles}
The models specified in the previous subsection are strictly non-rotating, as the DF depends on the modulus of $J_\phi$. Here, we generalize the models to allow for rotation, using the procedure described in~\cite{Binney2014}.

To our even DF $f_e(\bs{J})$, we add an odd component $f_o(\bs{J})$ such that the full DF is given by
\begin{equation}
f(\bs{J}) = f_e(\bs{J})+\frac{k}{1-|k|}f_o(\bs{J}) \text{ where } f_o(\bs{J})=h(J_\phi)f_e(\bs{J}).
\end{equation}
The function $h(J_\phi)$ is odd and restricted to $-1\leq h(J_\phi) \leq 1$. The magnitude of the rotation is governed by the parameter $|k|$ which varies from $0$ for the non-rotating case to $\tfrac{1}{2}$ for the maximally rotating case. Negative $k$ produces models that rotate in the opposite direction.

Rotation is an antisymmetric property in action space as in velocity space. The introduction of rotation as a function of the actions does not contribute to the density of the model, which depends solely on the even part of the distribution function $f_e$. Rotation can therefore be switched on or off for our action-based models with little computational cost, as it does not interfere with self-consistency.

We work with two simple analytic rotation profiles:
\begin{equation}
\begin{split}
h_T(\bs{J})&=\tanh\Big(\frac{J_\phi}{\chi}\Big),\\
h_E(\bs{J})&=\frac{\sqrt{2\e}}{\chi}J_\phi\exp\Big(-\frac{J_\phi^2}{\chi^2}\Big).\\
\end{split}
\label{Eqn::RotProfiles}
\end{equation}
Both models have a single parameter $\chi>0$ which controls the slope of $h$ at small $J_\phi$. The models differ in that $h_T$ tends to a fixed rotation at large radii, whereas $h_E$ produces zero rotation at large radii such that the rotation is localized to angular momenta $|J_\phi|\lesssim2\chi$. This reduces the number of orbits with high angular momentum relative to the $z$-axis, and thus changes the shape of the rotation curve at its peak amplitude, causing a faster decay in the mean projected velocity with radius. In Figure~\ref{Fig::sigma_overplot}, we show the numerical rotation curves $\langle v_{||}\rangle(b)$ produced by our self-consistent models, where $b$ is the distance from the $z$-axis in the plane of sight. We note that the properties of $h(J_\phi)$ largely translate into similar properties of $\langle v_{||}\rangle$, giving the exponential curves a generally higher amplitude and steeper gradient at small $\chi$. Over-plotted is the binned data for NGC 4372, for which the difference between these models is significant: a steeper gradient is required to fit the two data points with especially large radial velocities. We will elaborate further on the relative merits of the hyperbolic tangent and exponential models in Section~\ref{Sec::Discussion}.

\subsection{Model validation}
Here, we examine how successfully our models reproduce the observed velocity dispersion profiles before embarking on a full fitting of the data.

\begin{figure*}
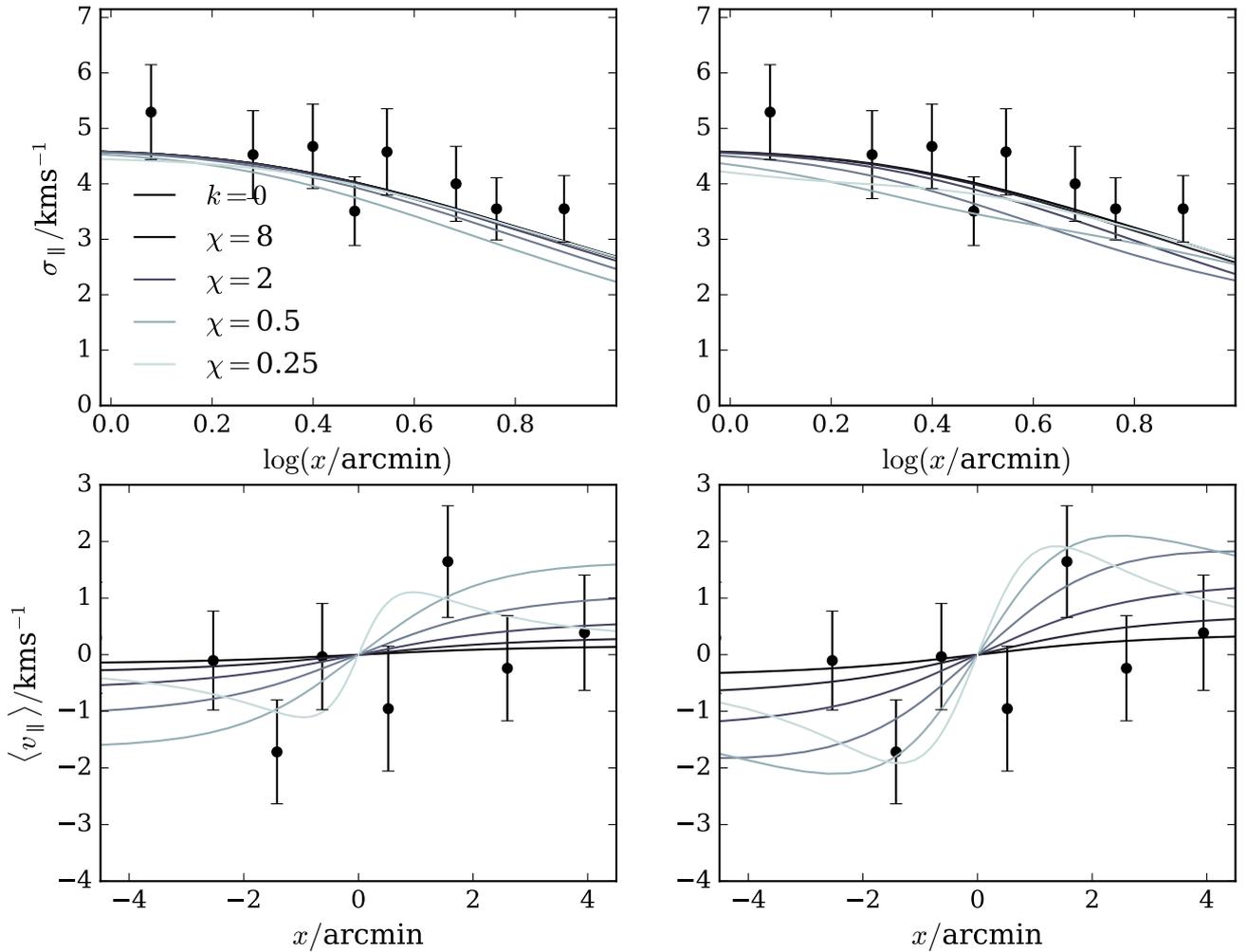

$$\includegraphics[width=\linewidth]{{{figs/Fig7}}}$$
\caption{Projected circularly-averaged velocity dispersion profiles (above) and rotation profiles (below) for those of our models with maximum rotation ($k=0.5$), over-plotted with binned data for NGC 4372, from the sources in Sec.~\ref{Sec::Data}. The velocity scale used for NGC 4372 is computed using the values in Table~\protect\ref{Table::FitResults} and the models have been scaled by the average Milky Way globular cluster mass-to-light ratio $1.98 \pm 0.19$ from~\protect\cite{McLaughlin2005}. The left-hand plots show results for the hyperbolic tangent rotation profile, while the right-hand plots show results for the exponential rotation profile. We see a promising agreement between the scales and overall shapes of the models and data, and note that the range of $\chi$ values between $\chi = 0.25$ and $\chi=8$ gives a reasonable spread of gradients about the data points, where higher values of $\chi$ are represented by darker lines.}
\label{Fig::sigma_overplot}
\end{figure*}

To obtain the velocity dispersion profiles plotted against the binned data in Figure~\ref{Fig::sigma_overplot}, we use the average V-band mass-to-light ratio for the Milky Way globular clusters from~\cite{McLaughlin2005}, $\Upsilon/(M_{\odot}/L_{\odot})=1.98 \pm 0.19$, where the error represents the standard deviation over all clusters. We see a promising agreement between the shape and scale of the data in the sample case of NGC 4372. We also note that in the case of maximum rotation ($k=0.5$), for which we expect the largest spread of gradients in the velocity dispersion profile, the range of $\chi$ values from $\chi=0.25$ to $\chi=8$ gives an appropriate spread of curves to match the data.

\section{Data analysis}\label{Sec::DataAnalysis}
With our model framework in place, we now discuss how we fit the models to the data. As each model calculation is expensive, we opt to compute the models on a grid of parameters and using a novel interpolation scheme we are able to accurately compute model properties at any parameter values. In Section~\ref{Section::ModelRepresentation}, we present the novel representation of our models that enables accurate interpolation. In Section~\ref{Section::ModelChoice}, we present our choices for the computed grid of models. Finally, in Section~\ref{Section::Fitting}, we give our formalism for fitting the models to the data.

\subsection{Compact model representation}\label{Section::ModelRepresentation}
For our analysis, we require the line-of-sight (l.o.s.) velocity $v_{||}$ distribution as a function of on-sky position $(x,y)$ for each set of parameters $\mathcal{P}$: $f(v_{||}|x,y,\mathcal{P})$. In general, the models presented in this paper are not fast to compute. It takes approximately 300 seconds to compute a self-consistent model to an error threshold of $0.1\%$, and a further 50 seconds to find each l.o.s. velocity distribution. This is not fast enough to use in an MCMC algorithm. However, as we anticipate the model properties vary slowly with the flattening and rotation parameters ($q_z,\chi,k$), we need only to interpolate over a small set of models that span the anticipated range in these parameters.

This then raises the question as to what properties of the models should be interpolated and whether there is an efficient representation of the models that could be employed. For the non-rotating models, we have found that the l.o.s. velocity distributions are very well fit by q-Gaussian profiles \citep{Tsallis2009} defined by
\begin{equation}
\mathcal{W}(y;q,\beta)\propto (1-\beta(1-q)x^2)^{1/(1-q)}.
\end{equation}
The q-Gaussians offer an improvement over a Gaussian (the limit of the q-Gaussian as $q\rightarrow1$) as for $q<1$ they fall to zero at finite $x$ and therefore follow the behaviour of the l.o.s. velocity distributions beyond the escape speed.

\begin{figure}
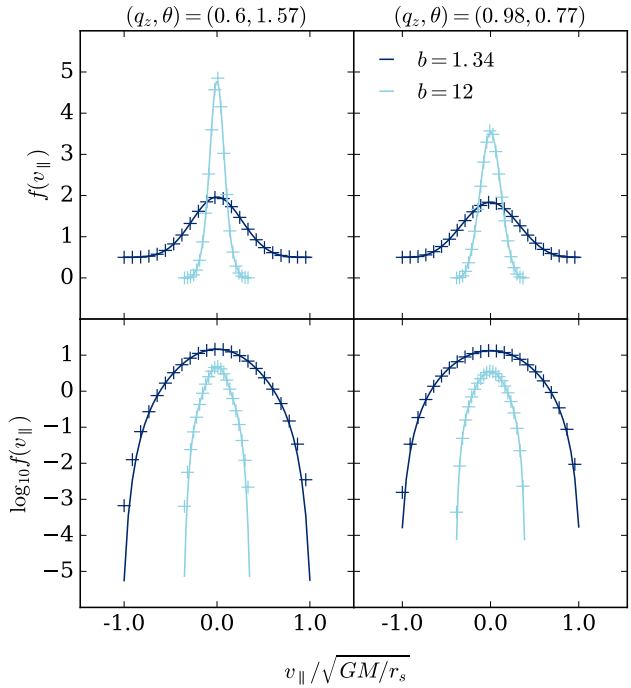

$$\includegraphics[width=\columnwidth]{{{figs/Fig8}}}$$
\caption{Linear and logarithmic plots of q-Gaussian fits to the l.o.s. velocity distributions at the two most extreme axial ratios $q_z$ represented in our parameter space. These are taken at a small angle to the z-axis ($\phi^{\prime}=1.47$) to emphasise the effect of flattening on the l.o.s. velocities. The fits are tested close to the scale radius ($b=1.34$) and at the largest impact parameter represented ($b=12$), demonstrating the relatively greater effect of the axial ratio at larger impact parameters. The crosses correspond to our models and the continuous lines to the fitted Gaussians, with all distributions normalised to unit area. The $b=1.34$ plots have been shifted by $+0.5$ on the linear scale and $\times 10$ on the logarithmic scale. Note that the q-Gaussians perform very well in both cases, over at least five orders of magnitude.}
\label{Fig::QGauss_NonRot}
\end{figure}

In Fig.~\ref{Fig::QGauss_NonRot}, we show two examples of l.o.s. velocity distributions for the non-rotating Modified Plummer model with the maximum and minimum flattening used in this work ($q_z=0.6$ and $q_z=0.98$) along with the best q-Gaussian fits. The q-Gaussian does a fantastic job of reproducing the distribution over about five orders of magnitude, for different models at different on-sky positions.

\begin{figure}
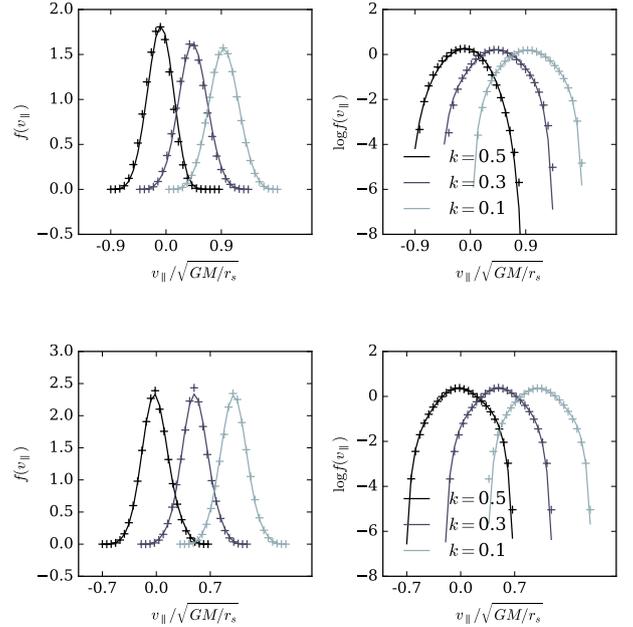

$$\includegraphics[width=\columnwidth]{{{figs/Fig9a}}}$$%
$$\includegraphics[width=\columnwidth]{{{figs/Fig9b}}}$$
\caption{Linear and logarithmic plots of q-Gaussian fits to the l.o.s. velocity distributions over the range of rotation parameters $k=0.1,0.3,0.5$ with $\chi=0.25$ and axial ratio $(q_z=0.6,\theta=0.77)$. Above we show the resulting plots for the hyperbolic tangent rotation curve, while below we show the plots for the exponential rotation curve. We have used a position $b=1.34,\phi^{\prime}=0.65$, at which the onset of rotation has a visible effect. The successive plots have been shifted horizontally by $0.5\sqrt{GM/r_s}$ for aesthetic purposes. Note the remarkable agreement for all rotation parameters over 7 orders of magnitude.}
\label{Fig::qgaussAllRot}
\end{figure}

Inspired by the Gauss-Hermite representation that has been widely employed in characterizing the line profiles of external galaxies \citep{vanderMarel1993}, we here use a novel approach to representing the l.o.s. velocity distribution using a q-Gaussian orthogonal polynomial basis expansion. In Appendix~\ref{Appendix::qGauss}, we describe how a general distribution can be represented by a set of q-Gaussian polynomials, give explicit expressions for the first five polynomials and describe how the parameters of the q-Gaussian weight function ($q,\beta$) are chosen. Each l.o.s. distribution can then be represented by eight numbers -- $q$, $\beta$, the five coefficients $\mathfrak{q}_i$ and the shift in position of the l.o.s. velocity maximum due to rotation, $v_0$. In Fig.~\ref{Fig::qgaussAllRot}, we show the close agreement between our models and the q-Gaussian fits over the range of rotation parameters $k$, and in Appendix~\ref{Appendix::qGauss}, we demonstrate the representation's success in reproducing l.o.s.~distributions over the entire parameter grid, via interpolation. A close agreement can be seen between profiles interpolated from the q-Gaussian parameters and profiles computed directly via self-consistency.

\subsection{Choice of models}\label{Section::ModelChoice}
Our models have three physical parameters, the viewing angle $\theta$ to the $z$-axis in spherical co-ordinates, the amplitude of rotation $k$ and the angular momentum scale $\chi$. Axisymmetry of the models gives a natural range of $\theta$ between $\theta = 0$ and $\theta = \pi/2$, and for each cluster we choose a lower limit for $\theta$ by setting the maximum intrinsic cluster ellipticity according to the maximum ellipticity $\epsilon = 0.27$ reported for any globular cluster in~\citet[2010 edition]{Harris2010}. Using the relation between observed ($q_{\text{obs}}$) and intrinsic ($q_{\text{isc}}$) axial ratios given by
\begin{equation}
q_{\text{obs}}^2 = q_{\text{isc}}^2 \sin^2{\theta} + \cos^2{\theta},
\end{equation}
the values of $q_{\text{obs}}$ from~\cite{WhiteShawl1987} and the lower limit $q_{\text{isc}}=0.73$ from~\citet[2010 edition]{Harris2010} give a lower limit on $\theta$ for each cluster.
The analytical form of the rotation curves gives a range of $k$ between $k=-0.5$ and $k=0.5$. We have determined an appropriate range for $\chi$ by comparing the projected mean velocity curves for the maximally-rotating case ($k=\pm 0.5$) to the velocity dispersion data as shown in Fig.~\ref{Fig::sigma_overplot}. The range of possible gradients for $\sigma/\sqrt{GM/r_s}$ is largest for maximal $k$ and so we wish to ensure that even for this extreme case, we do not obtain unreasonable results. We find that a range $\chi/\sqrt{GMr_s}=0.25,0.5,1,2,4$ and $8$ gives a reasonable spread of values. 

Given a set of model parameters $(\theta,k,\chi)$, we must evaluate the model at range of on-sky positions. We use a on-sky polar coordinate with radius $b$ and polar angle $\phi^\prime$. We use a logarithmic grid of $b$ values between $b=0.5\arcmin$ and $b=12\arcmin$, matching the range of the spectroscopic data from Sec.~\ref{Sec::spectroscopic_data}. Axisymmetry of the models gives a natural range of $\phi^{\prime}$ between $\phi^{\prime}=0$ and $\phi^{\prime}=\pi/2$. We evaluate the q-Gaussian expansion parameters for all our models such that for a given vector $(\theta,k,\chi,b,\phi^\prime)$, any model line-of-sight velocity distribution at any position can be evaluated via linear interpolation on the q-Gaussian parameters.

\subsection{Fitting models to data}\label{Section::Fitting}

\begin{table*}
\caption{Results of fitting the dynamical models to the data: three different models are considered -- norot is a non-rotating model whilst tanh and exp correspond to rotating models with the form given in equation~\protect\eqref{Eqn::RotProfiles}. The rotating models are described by the amplitude and scale parameters, $k$ and $\chi$ respectively. $\theta$ is the angle of inclination to the rotation axis. $v_\mathrm{sys}$ is the systemic velocity and $\Upsilon$ the mass-to-light ratio. $\Delta_i$ gives the systematic offsets between different sets of spectroscopic data. Finally, $K$ gives the Bayesian factor for the rotating models compared to the non-rotating model.}
\input{Table2.dat}
\label{Table::Results}
\end{table*}

We fit the model parameters $\mathcal{P}$ using the data $D$ by evaluating the posterior
\begin{equation}
p(\mathcal{P}|D)\propto \mathcal{L}(D|\mathcal{P}) p(\mathcal{P}).
\end{equation}
We define $\mathcal{P}=(\theta,k,\chi,v_\mathrm{sys},\Upsilon)$ where $v_\mathrm{sys}$ is the systemic velocity of the entire cluster and $\Upsilon$ the mass-to-light ratio.

Our data $D$ is a set of radial velocities $\{v_{||,i}\}$ and associated uncertainties $\{\sigma_{||,i}\}$ at a set of positions $\{x_i,y_i\}$ defined with respect to a coordinate system aligned with the position angle of the cluster \citep[taken from][]{WhiteShawl1987}. We work with the likelihood
\begin{equation}
\mathcal{L}(D|\mathcal{P})\propto \prod_i\int\dd v'_{||}\, p(v_{||,i}|v'_{||},\sigma_{||,i})p(v'_{||}|x_i,y_i,\mathcal{P}).
\end{equation}
The uncertainty distribution $p(v_{||,i}|v'_{||,i},\sigma_{||,i})$ is a normal distribution $\mathcal{N}(x;\mu,\sigma)$ such that the integral over the uncertainties can be performed by Monte Carlo integration using $N$ samples $\{v'_{||,i,n}\}$ from the distribution $\mathcal{N}(v'_{||};v_{||,i},\sigma_{||,i})$. The likelihood is then given by
\begin{equation}
\mathcal{L}(D|\mathcal{P})\propto \prod_i\frac{1}{N}\sum_{n=1}^{n=N} p(v'_{||,i,n}|x_i,y_i,\mathcal{P}).
\label{Eqn::MCsumLikelihood}
\end{equation}
We opt to work with an outlier model such that
\begin{equation}
p(v_{||}|x,y,\mathcal{P}) = (1-\epsilon)f(v_{||}|x,y,\mathcal{P})+\epsilon f_\mathrm{out}(v_{||}|x,y,\mathcal{P}),
\end{equation}
where $f$ is the likelihood of our dynamical model (computed via the q-Gaussian expansion) and $f_\mathrm{out}$ is the likelihood of the outlier model. We model $f_\mathrm{out}$  as $\mathcal{N}(v_{||};\mu_\mathrm{out},\sigma_\mathrm{out})$.

One further complication is the potential combination of $N_d$ different datasets for which we add $N_d$ additional \emph{offset} parameters $\Delta_k$ (with $\Delta_1=0$) such that equation~\eqref{Eqn::MCsumLikelihood} reads
\begin{equation}
\mathcal{L}(D|\mathcal{P})\propto \prod_{k=1}^{k=N_d}\prod_i\sum_{n=1}^{n=N} p(v'_{||,i,n}+\Delta_k|x_i,y_i,\mathcal{P}),
\label{Eqn::MCsumLikelihoodOffset}
\end{equation}
where the product $i$ runs over the data from data source $k$. These parameters account for possible zero-point systematic variations between different radial velocity sources.

In total, we have $(7+N_d)$ parameters in our fits. The fitting of parameters is performed in two steps. First, we maximise the posterior using Powell's method. We perform this procedure 5 times from a set of initial parameters drawn from a broad Gaussian. We choose the highest posterior result from the set of optimizations and seed a set of walkers about a much narrower Gaussian that explore the posterior using the \emph{emcee} algorithm \citep{emcee} using $60$ walkers with a burn-in and production of $\sim2000$ and $\sim1000$ steps respectively. We set the number of samples from the radial velocity uncertainties at $N=500$ and re-runs with $N=2500$ produced negligible changes in the results. We adopt uniform priors in the parameters $\cos\theta$, $k$, $\mu_\mathrm{out}$, $v_\mathrm{sc}$ and $\Delta_i$. We adopt uniform priors in the logarithm of the parameters $\chi$, $\epsilon$ and $\sigma_\mathrm{out}$. For $\Upsilon$, we use a Gaussian prior $\mathcal{N}(1.98,0.19)$ taken from the ensemble properties of the globular clusters from \cite{McLaughlin2005}.

\subsection{Results}

\begin{figure*}
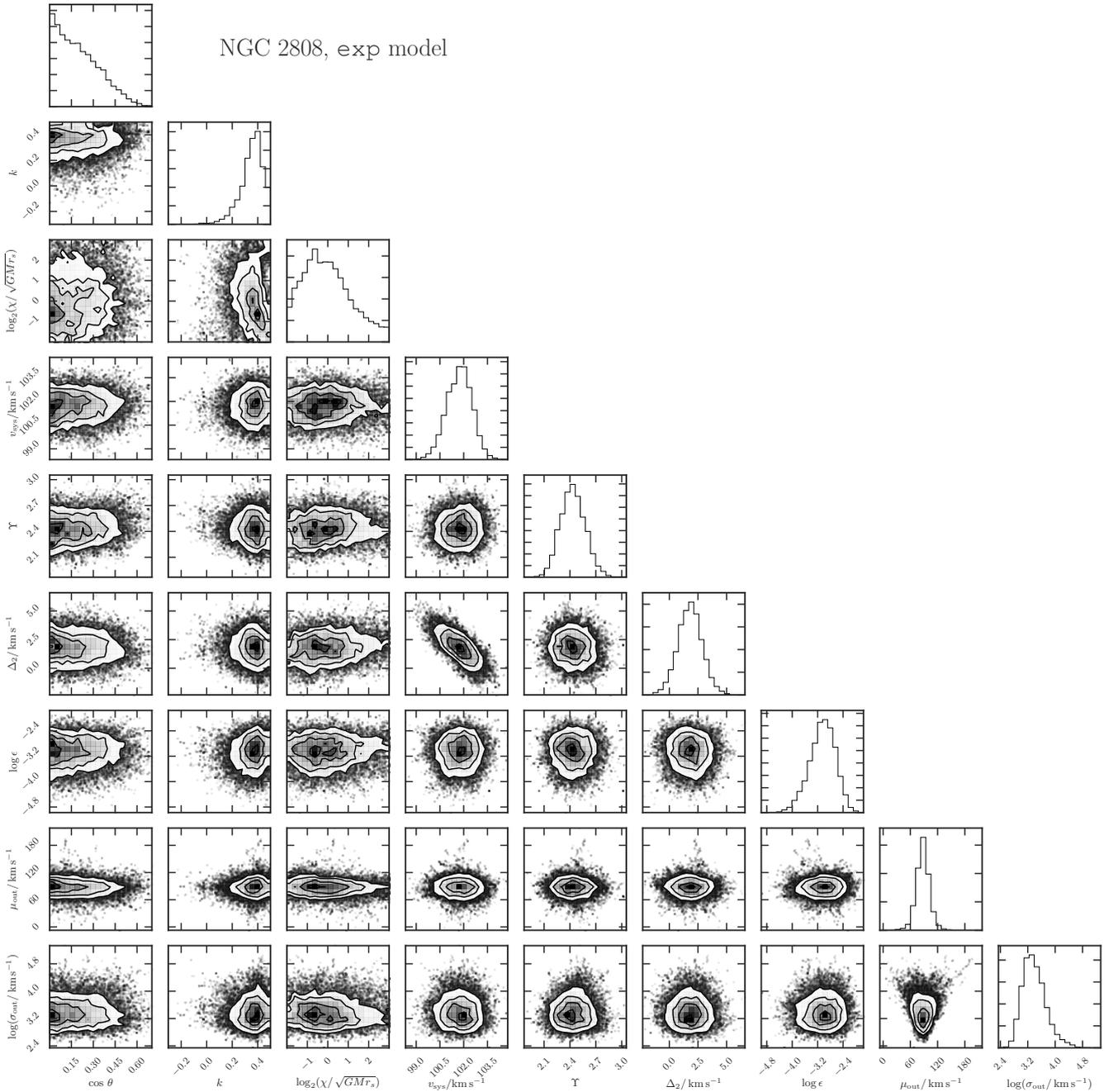

$$\includegraphics[width=\textwidth]{{{figs/Fig10}}}$$
\caption{Corner plot for the exponential rotating model fit to the data from NGC 2808. $\theta$ is the inclination angle, $k$ and $\chi$ the rotation profile amplitude and scale, $v_\mathrm{sys}$ the systemic velocity, $\Upsilon$ the mass-to-light ratio, $\Delta_2$ the offset between the systemic velocities of the two data subsamples used and $\epsilon$, $\mu_\mathrm{out}$ and $\sigma_\mathrm{out}$ are the parameters describing the outlier model. The four contour levels enclose $\sim12$, $\sim39$, $\sim68$ and $\sim86\,\percent$ of the probability. We have used the package from \protect\cite{corner} to produce this plot.}
\label{Figure::ExampleCornerPlot}
\end{figure*}

In Figure~\ref{Figure::ExampleCornerPlot}, we show the results of one of the MCMC runs fitting the exponential model to the data from NGC 2808. This corner plot displays the generic features seen for all the clusters but is chosen because NGC 2808 is the cluster in our sample that has the most convincing evidence for rotation. In general, we see a lack of correlations in our parameters except between the systemic velocity $v_\mathrm{sys}$ and the $\Delta_i$. Anti-correlations are expected between these values, as for each data subsample the constraint is on $v_\mathrm{sys}+\Delta_i$.

\begin{figure*}
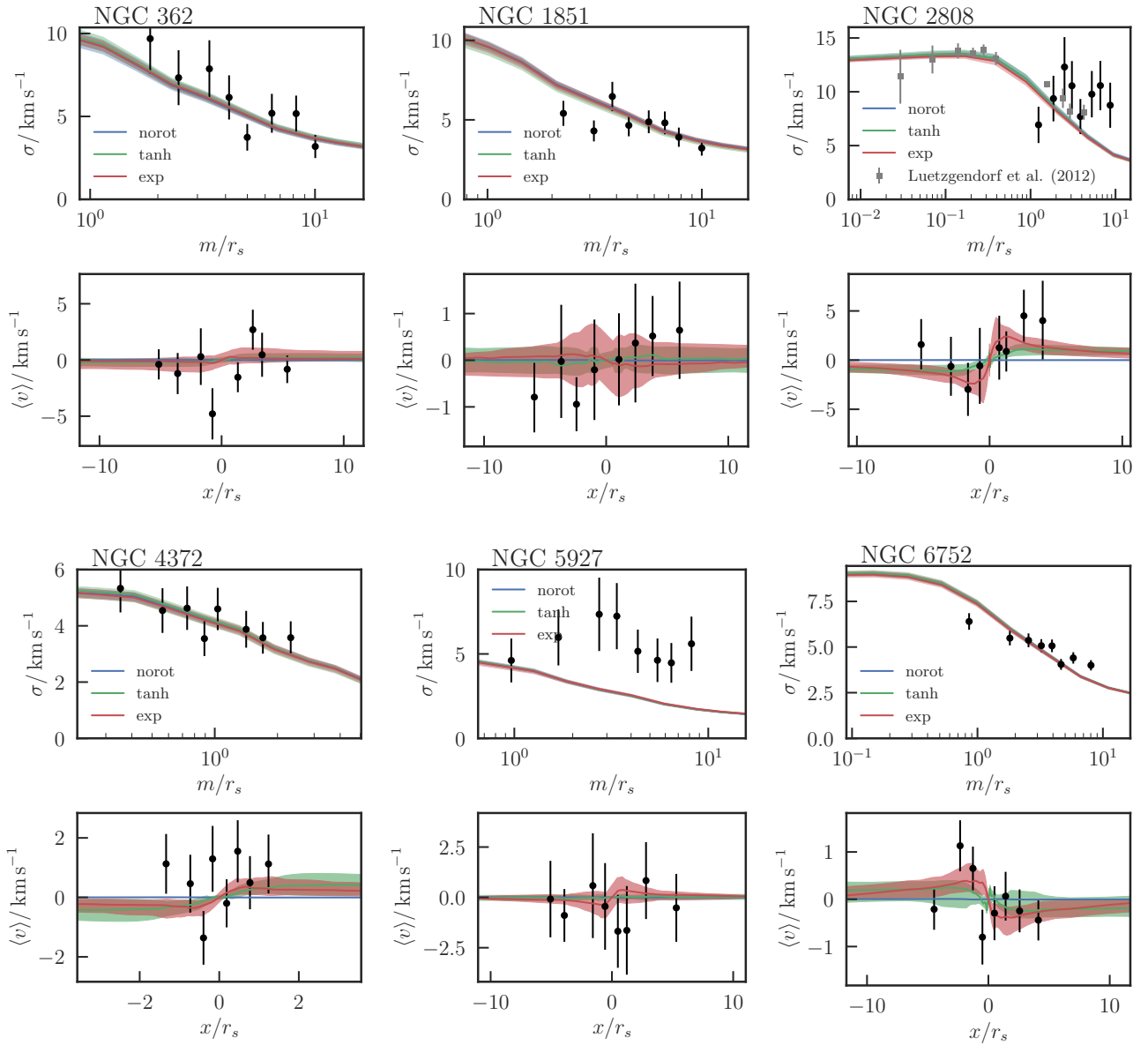

\begin{minipage}{.32\textwidth}
$$\includegraphics[width=\columnwidth]{{{figs/Fig11a}}}$$
\end{minipage}
\begin{minipage}{.32\textwidth}
$$\includegraphics[width=\columnwidth]{{{figs/Fig11b}}}$$
\end{minipage}
\begin{minipage}{.32\textwidth}
$$\includegraphics[width=\columnwidth]{{{figs/Fig11c}}}$$
\end{minipage}
\begin{minipage}{.32\textwidth}
$$\includegraphics[width=\columnwidth]{{{figs/Fig11d}}}$$
\end{minipage}
\begin{minipage}{.32\textwidth}
$$\includegraphics[width=\columnwidth]{{{figs/Fig11e}}}$$
\end{minipage}
\begin{minipage}{.32\textwidth}
$$\includegraphics[width=\columnwidth]{{{figs/Fig11f}}}$$
\end{minipage}
\caption{Dispersion and mean velocity profiles fit results: for our six globular clusters we show the line-of-sight dispersion profiles $\sigma$ as a function of elliptical radius $m=\sqrt{x^2+y^2/q^2}$ and the mean velocity $\langle v\rangle$ along the major axis. The black points are computed using a Gaussian mixture model on the data split into eight equally-populated bins. The lines and coloured bands show the median and one-sigma confidence region computed from the models using samples from our MCMC chains (note for the mean velocity we bin in $x$ but the models are computed along $y=0$). Blue corresponds to the non-rotating model, whilst green and red correspond to rotating models with hyperbolic and exponential rotation curves as described in the text. For NGC 2808, we also show the data from \protect\cite{Luetzgendorf2012} in grey, which was not used in the fits.}
\label{Figure::ResultsProfiles}
\end{figure*}

In Table~\ref{Table::Results}, we present the median values for each of the parameters in the three models for each cluster, along with their $68\,\percent$ confidence intervals. The systemic velocities, mass-to-light ratios and velocity offset parameters are all well constrained and take consistent values for all three models. We will discuss these in the next section. In general, we find that the data favour lower values of $\cos\theta$ such that very small inclination angles are ruled out. These models have the most extreme flattening consistent with the observed ellipticity, and consequently rule out high degrees of flattening along the line of sight. For the tanh model, we measure $k$ consistent with zero at the $\sim1\sigma$ level except in the case of NGC 2808. The rotation curve scale $\chi$ is more poorly constrained than the other parameters, with essentially the entire allowed range being consistent with the data. For the exp model, we find that $k$ is non-zero at the $1\sigma$ level for NGC 2808 and NGC 6752 (and marginally for NGC 5927). Again, $\chi$ is poorly constrained but there is a tendency for $\chi\approx1-2$ such that the data favours a rotating core within radius $r_s$.

In Figure~\ref{Figure::ResultsProfiles}, we show the line-of-sight velocity dispersion and mean velocities of the models compared to the data. The velocity dispersion is computed in elliptical bins, with ellipticity equal to the observed ellipticity from Table~\ref{Table::Summary} for both the models and the data. The mean velocities are computed by binning the data along the major axis, whilst for the models we simply compute the mean velocity along the major axis. We see that the dispersions for all three sets of models are very similar and agree well with the data. In the cases of NGC 5927 and NGC 2808, the models appear to be a poor representation of the data, but this can partly be attributed to the presence of outliers (which our models find are significant in these clusters). These cases are discussed further in Sections~\ref{Sec::NGC2808} and~\ref{Sec::NGC5927}. We see that for most clusters the data points show a low level of rotation, which is replicated by the models. However, in NGC 2808 and NGC 6752, the data appear to produce a rotation signal ($k>0$ at the $1\sigma$ level). In NGC 2808, this signal is fitted nicely by both the tanh and exp models, but for NGC 6752 it appears only the exp model can fit the rotation profile.

\begin{figure*}
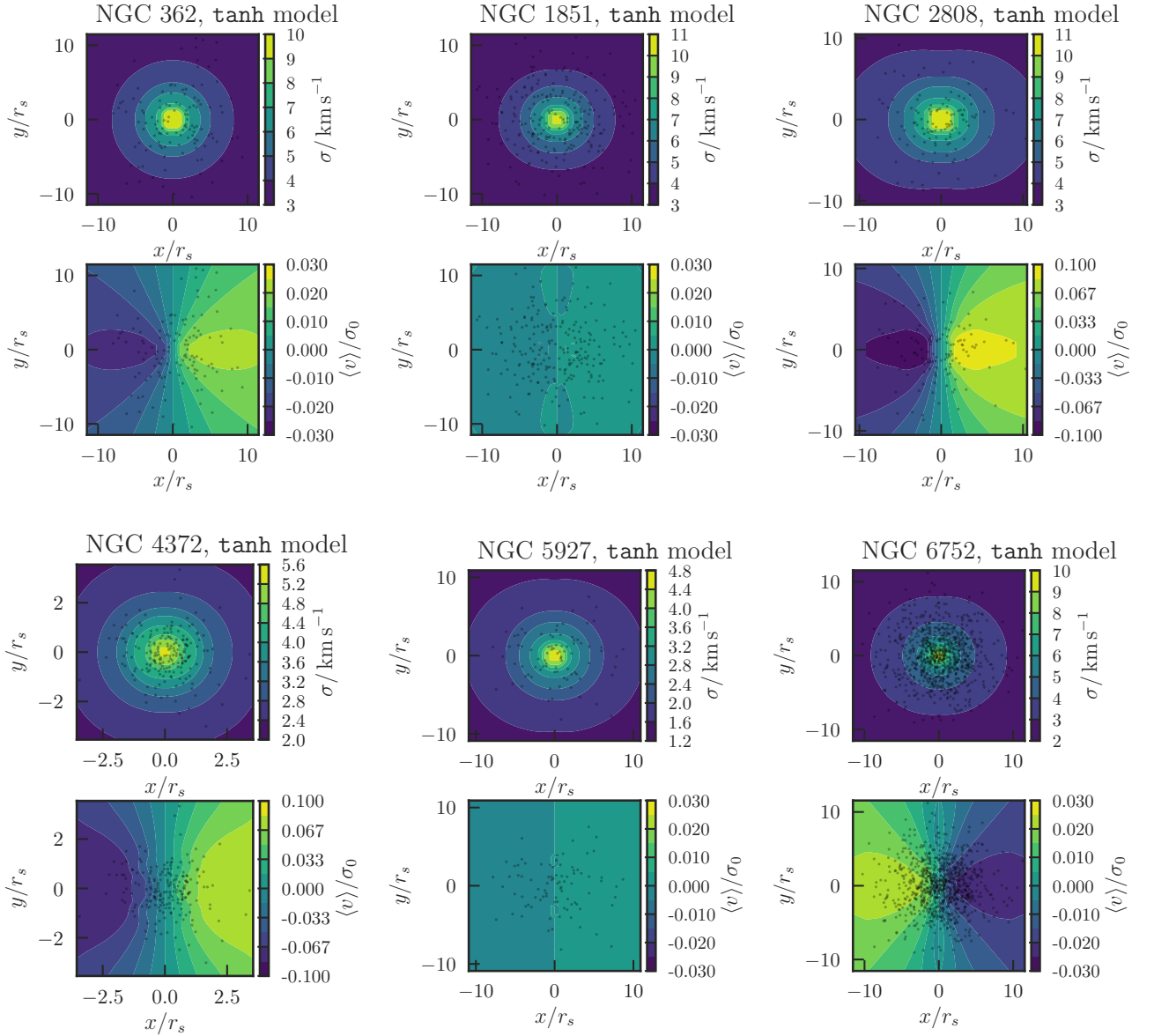

\begin{minipage}{.32\textwidth}
$$\includegraphics[width=\columnwidth]{{{figs/Fig12a}}}$$
\end{minipage}
\begin{minipage}{.32\textwidth}
$$\includegraphics[width=\columnwidth]{{{figs/Fig12b}}}$$
\end{minipage}
\begin{minipage}{.32\textwidth}
$$\includegraphics[width=\columnwidth]{{{figs/Fig12c}}}$$
\end{minipage}
\begin{minipage}{.32\textwidth}
$$\includegraphics[width=\columnwidth]{{{figs/Fig12d}}}$$
\end{minipage}
\begin{minipage}{.32\textwidth}
$$\includegraphics[width=\columnwidth]{{{figs/Fig12e}}}$$
\end{minipage}
\begin{minipage}{.32\textwidth}
$$\includegraphics[width=\columnwidth]{{{figs/Fig12f}}}$$
\end{minipage}
\caption{2D hyperbolic rotation curve model fits: each set of panels shows the line-of-sight velocity dispersion on the sky and the mean line-of-sight velocity scaled by the central dispersion. On each panel, we also display the location of the spectroscopic data in faint black. Note the different ranges used for $\langle v\rangle/\sigma_0$ (the minimum range is $\pm3\,\percent$).}
\label{Fig::Tanh2D}
\end{figure*}

\begin{figure*}
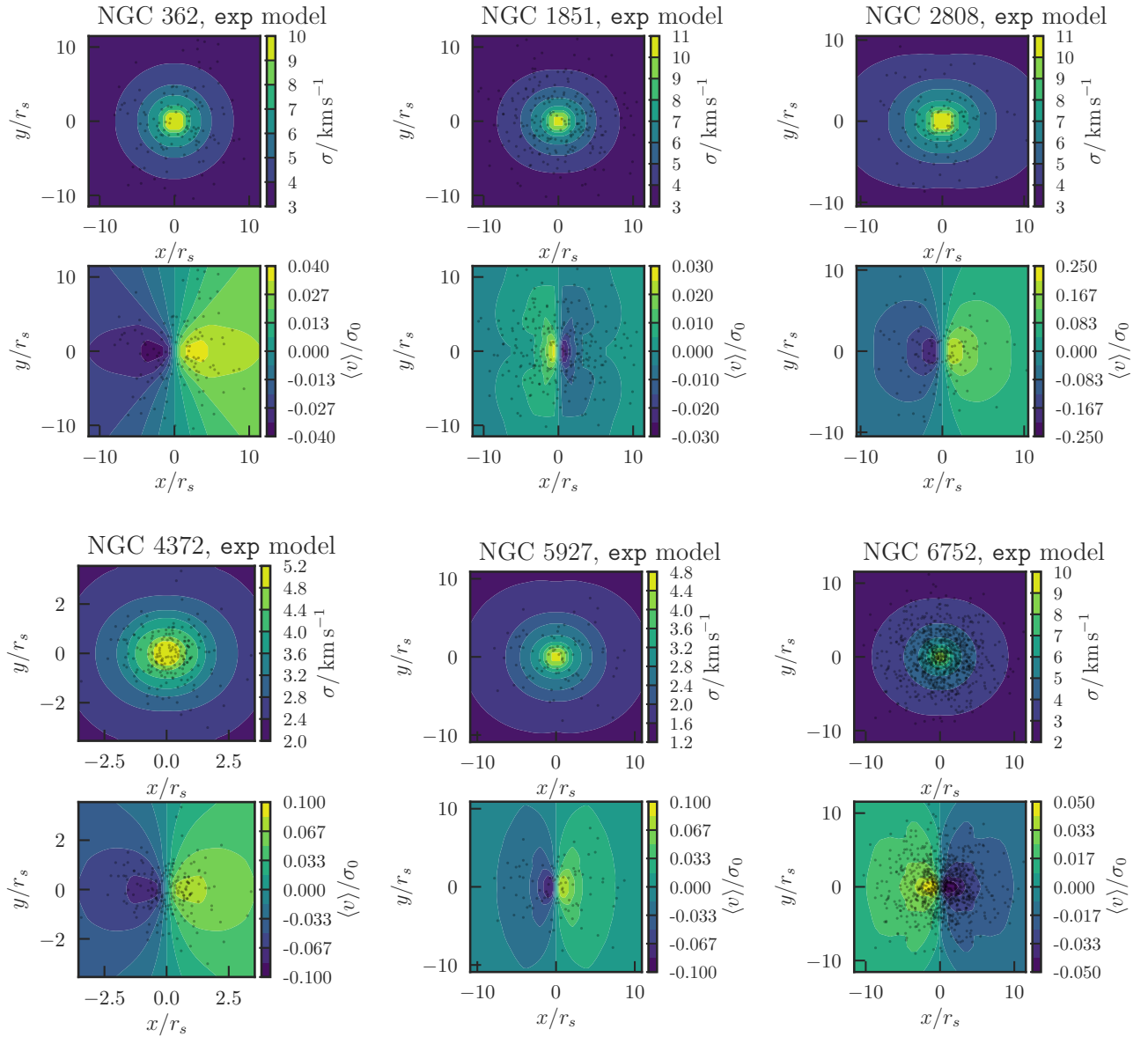

\begin{minipage}{.32\textwidth}
$$\includegraphics[width=\columnwidth]{{{figs/Fig13a}}}$$
\end{minipage}
\begin{minipage}{.32\textwidth}
$$\includegraphics[width=\columnwidth]{{{figs/Fig13b}}}$$
\end{minipage}
\begin{minipage}{.32\textwidth}
$$\includegraphics[width=\columnwidth]{{{figs/Fig13c}}}$$
\end{minipage}
\begin{minipage}{.32\textwidth}
$$\includegraphics[width=\columnwidth]{{{figs/Fig13d}}}$$
\end{minipage}
\begin{minipage}{.32\textwidth}
$$\includegraphics[width=\columnwidth]{{{figs/Fig13e}}}$$
\end{minipage}
\begin{minipage}{.32\textwidth}
$$\includegraphics[width=\columnwidth]{{{figs/Fig13f}}}$$
\end{minipage}
\caption{2D exponential rotation curve model fits: see Fig.~\ref{Fig::Tanh2D} for details.}
\label{Fig::Exp2D}
\end{figure*}

In Figures~\ref{Fig::Tanh2D} and~\ref{Fig::Exp2D}, we show the median 2D line-of-sight velocity dispersions and mean velocities for the tanh and exp models with the data points overplotted. We see that the 2D line-of-sight dispersions are very similar for both models. We also note the difference in the shape of the mean velocity surfaces, with the exp model producing a more pronounced rotating core than the tanh models. We see that in the cases of NGC 2808, NGC 6752 and NGC 5927 a more central rotating core is favoured, whilst for NGC 362 and NGC 4372, a more extended rotation curve is required.

\section{Discussion}\label{Sec::Discussion}

\subsection{Evidence for rotation}
\label{Sec::Rotation}

Although the values of $k$ and $\chi$ computed for our models point towards the presence of rotation in a few cases, the evidence for rotation appears quite weak. To quantify this, we have computed the evidences for both the hyperbolic tangent and exponential rotating models, relative to the evidence for the non-rotating model, given by
\begin{equation}
K=\frac{Z_\text{rot}}{Z_\text{no rot}} = \frac{\int\mathrm{d}\mathcal{P}_\text{rot}\,p(\mathcal{P}_\text{rot}|D)}{\int\mathrm{d}\mathcal{P}_\text{no rot}\,p(\mathcal{P}_\text{no rot}|D)}.
\end{equation}
We evaluate the evidence for each model using the Multinest algorithm \citep{Multinest} via a Python interface \citep{PyMultinest}. The results are reported in Table~\ref{Table::Results} along with an estimate of the uncertainty output by the algorithm. Following the guidelines from \cite{BayesFactor}, we interpret values of $2\ln K>2$ as positive evidence for cluster rotation. We see that when using the tanh model, the evidence for rotation is deemed strong only for NGC 2808. NGC 4372 has a weak indication of rotation ($2\ln K\sim 1.5$) but all other clusters have evidence ratios $K<1$ indicating clearly that the introduction of rotation in the form of a flat hyperbolic rotation curve does not significantly improve the model fits. For the exponential rotation model, the evidence ratio increases over the hyperbolic model for NGC 2808, NGC 4372 and NGC 5927, indicating that the rotation in these clusters is more centralised. NGC 2808 displays a convincing level of rotation, while the other two clusters display more modest evidence that rotation is present ($2\ln K\approx2$). For NGC 362, NGC 1851 and NGC 6752, the Bayes factors for the exponential and hyperbolic tangent models are essentially unity, indicating that the evidence for rotation is very weak, if rotation is present at all.

Our rotation parameters $k$ and $\chi$ can be used to calculate the physical rotation parameter $A_{\text{rot}}/\sigma_0$, a common measure for the degree of rotation in globular clusters~\citep[e.g.][]{Kacharov2014}, where $\sigma_0$ is the central velocity dispersion and $A_{\text{rot}}$ refers to the maximum amplitude of the projected mean velocity profile\footnote{We compute $A_\mathrm{rot}$ by finding the maximum mean velocity along the major axis for each set of parameters from our Monte Carlo chains and the position angles described in Section~\ref{Sec::Ellipticity}, and de-projecting by dividing by $\sin\theta$. Formally, we should compute the signal observed when viewing edge-on, but the discrepancy is small.}. In Figure~\ref{Fig::AvsE}, we plot this parameter against the intrinsic ellipticity $\epsilon=1-q_\mathrm{isc}$ (filled symbols), along with the corresponding values from the literature (empty symbols). We also show data from \cite{Bellazzini2012} and \cite{Bianchini2013} for a larger sample of globular clusters. We see that our clusters sit on the approximate positive correlation between ellipticity and rotation strength (that is also exhibited in the literature data), although there is a large degree of scatter. We have also shown the relationship between ellipticity and rotation strength for an isotropic spheroidal rotator from \cite{Binney2005} along with two lines for vertically anisotropic models (quantified by $\delta=1-\langle\sigma_{zz}^2\rangle/\langle\sigma_{xx}^2\rangle$). When a galaxy is flattened, the kinetic energy associated with vertical motion decreases relative to that associated with the planar motion. In order to maintain isotropy, some planar kinetic energy must be supplied by rotation. Below the isotropic rotator curve, $\langle\sigma_{zz}^2\rangle<\langle\sigma_{xx}^2\rangle$ as less of the planar motion is provided by rotation and must be compensated for by random motions. Therefore, beneath the isotropic rotator line we link the flattening more strongly to anisotropy. Above the line $\langle\sigma_{zz}^2\rangle>\langle\sigma_{xx}^2\rangle$ so we require more rotation to compensate for the decrease in random planar motions. We see that in general our clusters lie beneath the isotropic rotator line, indicating that the flattening of the clusters is related to anisotropy as opposed to rotation. However, for several clusters (NGC 362, NGC 5927 and NGC 2808), the rotation strength is still consistent with the isotropic rotator, suggesting that the flattening may be related to rotation. We note that our models are flexible enough to allow any combination of $A_\mathrm{rot}/\sigma_0$ and $\epsilon$ in this plane and that the observed distribution/correlation is not due to the models being too restrictive.

In Figure~\ref{Fig::AvsE}, we also show existing estimates from the literature for some of our clusters, which we will discuss along with the evidence for rotation in each case. Both NGC 362 and NGC 6752 have low evidence for rotation and we do not comment in any further detail on these clusters.

\subsubsection{NGC 1851}
As shown in Table~\ref{Table::Results}, we obtain negligible evidence for rotation in the case of NGC 1851, relative to the non-rotating case. NGC 1851 has been studied in detail by~\cite{Scarpa2011} using a sample of 184 radial velocities, comparable with our 221. In this work a maximum rotation amplitude of $A_{\text{rot}}<0.8\text{kms}^{-1}$ is quoted along with a central velocity dispersion $\sigma_0=10.4\text{kms}^{-1}$, in agreement with the value from~\citet[2010 edition]{Harris2010}. Given that this constitutes an upper limit on the rotation signal, we do not believe that our result is in conflict with the literature, but rather provides stronger evidence that the degree of rotation in NGC 1851 is negligible. In Figure~\ref{Fig::AvsE}, we see that our results agree with this literature value, to within error.

\subsubsection{NGC 2808}
\label{Sec::NGC2808}
NGC 2808 contains at least five different stellar populations with different radial distributions~\citep{Simioni2016}, making it a particularly complex cluster both to model and to observe. This is reflected in the fact that our model can provide a very good match to the integral-field spectroscopy and Fabry-Perot measurements of velocity dispersion obtained by~\cite{Luetzgendorf2012} for the inner regions of the cluster (grey points in Figure~\ref{Figure::ResultsProfiles}), but a much worse match to data in the outer regions. Despite this, Figure~\ref{Figure::ExampleCornerPlot} demonstrates that the parameters for this cluster are well-constrained in our models, such that we can draw conclusive evidence for rotation in this case.

\subsubsection{NGC 4372}
In the case of NGC 4372 we compare our results to those of~\cite{Kacharov2014}. Figure~\ref{Fig::AvsE} demonstrates that these values agree to within error, with ours having a significantly lower value of the rotation parameter. This may partially be explained by the fact that we do not group the radial velocity data into overlapping bins in the plane of the sky as in~\cite{Kacharov2014}, a procedure which potentially enhances the rotation signal.

As our results are in tension with those of \cite{Kacharov2014}, we have opted to also fit our models to the same dataset used by \cite{Kacharov2014}. Fitting the non-rotating model to the data gives a systemic velocity of $(75.8\pm0.4)\mathrm{km\,s}^{-1}$ and a mass to light ratio of $1.87\pm0.17$. These values agree well with \cite{Kacharov2014}, who find $(75.91\pm0.38)\mathrm{km\,s}^{-1}$ and $1.7\pm0.4$ (where we have quoted the model fit for an assumed inclination angle of $i=45\deg$ and using stars with $V<20$). The fact we have used a fixed density profile for our kinematic fits to the data is reflected in the smaller error bar for the mass-to-light ratio than the \cite{Kacharov2014} results. When fitting the exponential rotation law, we find the models favour a higher degree of rotation with $k=0.35^{+0.11}_{-0.23}$ (compared to $k=0.23\pm0.15$ from Table~\ref{Table::Results}), which corresponds to $A_\mathrm{rot}/\sigma_0=0.16^{+0.17}_{-0.11}$. This is lower than, but still consistent with, the value from \cite{Kacharov2014} of $A_\mathrm{rot}/\sigma_0=0.26\pm0.07$. We find the ratio of the exponential to the non-rotating model is $2\ln K=2.8\pm0.2$, which suggests the rotating model provides a slightly better fit than the non-rotating model, but that the data only marginally favour rotation.

\subsubsection{NGC 5927}
\label{Sec::NGC5927}
This cluster appears to have significant evidence for rotation in the case of the exponential model, but the most negligible rotation signal according to the hyperbolic tangent model. Although the exponential profile generally does a better job of fitting the rotation curve, such a disparity indicates that the data for NGC 5927 do not allow a robust conclusion to be drawn with regards to the rotation signal. Indeed, this cluster has the fewest radial velocity members in our sample, a total of 87. Furthermore, the surface brightness profile for NGC 5927 has a peculiar increase in density at around 10 arcseconds (possibly due to differential reddening) and so is not well-represented by our Modified Plummer model, despite being post-core-collapse~\citep{McLaughlin2005}. This is also noted by~\cite{Watkins2015}, who find that the unusual shape of the surface brightness profile gives rise to difficulties in fitting the spectroscopic data points above 1.5 arcminutes. We encounter a similar problem, whereby our model appears to give a good match to the shape of the velocity dispersion profile, but like the fits in~\cite{Watkins2015}, falls below the scale of these data points. We therefore stress that, despite the apparent evidence for rotation, our results are inconclusive in this case.

\begin{figure*}
$$\includegraphics[width=\textwidth]{{{figs/Fig14}}}$$
\caption{Relationship between intrinsic ellipticity and rotation for those clusters in our sample with evidence for rotation, NGC 2808 and NGC 5927. Empty symbols represent literature values~\protect\citep{Scarpa2011,Bellazzini2012,Bianchini2013,Kacharov2014}, while filled symbols represent our values. Black points represent a sample of clusters from the literature~\protect\citep{Bellazzini2012,Bianchini2013}. Note that our values are intrinsic properties whilst the literature data are the observed properties. The lines represent the upper limit for a maximally-rotating, self-gravitating sphere~\protect\citep{Binney2005}, where the solid line represents the isotropic case ($\delta=0$), and the dashed and dotted lines represent an increasing degree of radial anisotropy ($\delta=0.05$ and $\delta=0.1$ respectively).}
\label{Fig::AvsE}
\end{figure*}

\subsection{Mass-to-light ratios}
In Table~\ref{Table::MtoL}, we give our mass-to-light ratios along with mass-to-light ratios in the literature for the clusters in our sample. We see that our results are in good agreement with the literature values for all cases except for NGC 2808 (inconsistent with all other values at $>2\sigma$), which we have discussed in Section~\ref{Sec::NGC2808}. We also see a good agreement between the values of $\Upsilon$ for our three different rotation curves, indicating that these results are robust. An interesting comparison to our results is with the N-body simulations from \cite{Baumgardt2017}. As noted by this author, among others, mass segregation leads to underestimates of the dispersion from giant stars and thus to underestimates of the mass-to-light ratio. The mass-to-light ratios of \cite{Baumgardt2017} account for this effect. However, for only two of our clusters do we find that our estimates are smaller than those of \cite{Baumgardt2017} and in general the agreement is good.

\begin{table}
\caption{Comparison between our values of the mass-to-light ratio for each cluster in column (4) and the values from~\protect\cite{McLaughlin2005} in column (1),~\protect\cite{Watkins2015} in column (2) (from \protect\cite{Kacharov2014} in the case of NGC 4372) and \protect\cite{Baumgardt2017} in column (3).}
\input{Table3.dat}
\label{Table::MtoL}
\end{table}

\subsection{Velocity offsets}
In our analysis, we have allowed for arbitrary velocity offsets between different sources of data. In general, we find these velocity offsets are consistent with zero at the $2\sigma$ level. Comparing Figure~\ref{Fig::AvsE} and Table~\ref{Table::Results}, we see that the velocity offsets $\Delta_2$ and $\Delta_3$ between datasets are of the same order as the rotation signal. This is not a cause for concern, as the offset values are in good agreement between the non-rotating, hyperbolic tangent and exponential models. This indicates that our models have taken good account of offsets and have not erroneously interpreted them as rotation signals. As a cross-check of these offset measurements, we can compare them to the offsets measured using observations of the same stars in the different datasets. For the presented analysis, we naturally removed these. Table~\ref{Table::Offsets} gives the results for the offsets from the non-rotating model fits along with the median offsets measured from the duplicate observations. The Lardo sample has been calibrated so that duplicate offsets with the GES data are consistent with zero for all clusters but NGC 6752, for which the mean difference between duplicates is of magnitude $<1.0\,\mathrm{km\,s}^{-1}$. These offsets are replicated by our fits, particularly in the cases of NGC 1851 and NGC 4372, for which they are consistent with zero at the $1\sigma$ level. For NGC 2808 and NGC 5927 a larger positive offset, only consistent with zero at the $2\sigma$ level, is found. For NGC 6752 the two measures of offset between the Lardo and GES datasets agree that there is a small systematic discrepancy. For the comparison between the Kacharov and GES data, the picture from the two methods also consistently points to the Kacharov data moving systematically $0.5-1\,\mathrm{km\,s}^{-1}$ faster than the GES data (this is mirrored by the $1\,\mathrm{km\,s}^{-1}$ offset between the systemic velocities found using the full dataset and just the Kacharov data). Similarly for the Lane and GES data comparison, there is a small consistent, but not significant, positive offset. 

\begin{table}
\caption{Comparison between the offsets obtained from fitting the non-rotating model to each cluster compared to the offsets from duplicate observations between datasets. All quantities are given in $\mathrm{km}\,\mathrm{s}^{-1}$.}
\input{Table4.dat}
\label{Table::Offsets}
\end{table}

\subsection{Alternative models}

The models of \cite{VarriBertin2012} are similar to those presented here in that they are self-consistent rotating dynamical equilibria that are appropriate for the modelling of globular clusters. These authors consider two types of models: the first set of rigidly-rotating models depend on the Jacobi energy $E-\omega J_\phi$ whilst the second, more flexible, set are similar to truncated isotropic spherical models but depend on the argument
\begin{equation}
I = E-\frac{\omega J_\phi}{1+bJ_\phi^{2c}}.
\end{equation}
These models are differentially rotating with amplitude controlled by (a dimensionless version of) $\omega$, the scale $b$ and power $c$. As with our models, the moments of these models must be computed purely numerically. The authors note that these models have the nice properties of central isotropy whilst tending to tangential anisotropy in the outskirts with $\sigma^2_T/\sigma^2_R=2$. This should be compared to our models which tend to isotropy in the outskirts (although this can be simply adapted through modification of $D(\boldsymbol{J})$ (equation~\eqref{Eqn::DJ}) in the outskirts \citep[see][]{WilliamsEvans2015}. As with our models, the isodensity contours of the \cite{VarriBertin2012} models get rounder with radius. 

The models presented here are potentially much more flexible and powerful than those of \cite{VarriBertin2012}. For instance, by our procedure the density profile properties are separate from the rotation, whilst \cite{VarriBertin2012} find that the density profiles are altered when the rotation parameters are adjusted. In the most extreme cases, large degrees of rotation give rise to an increasing density with distance along the major axis -- a so-called torus structure -- whilst in our models the maximum rotation for a given density profile is mathematically constrained in an obvious way ($|k|<0.5$). Additionally, in our formulation, multiple components can be simply considered as the sum of action-based DFs whilst an equivalent procedure for DFs defined in terms of the energy is more awkward. Inspired by the work of \cite{VarriBertin2012}, it may be advantageous to construct truncated density profiles by following the same procedure as in Section~\ref{Sec::Model} and adopting a slightly more flexible rotation curve of
\begin{equation}
h(J_\phi) \propto J_\phi\exp\Big[-\Big(\frac{J_\phi}{\chi}\Big)^{2c}\Big]
\end{equation}
where we have introduced the parameter $c$ to allow for faster rotation profile decays at large radius. Throughout this paper, we have used $c=1$. We anticipate that the current data is entirely insensitive to modest changes in $c$.

\section{Conclusions}\label{Sec::Conclusions}

We have constructed axisymmetric rotating self-consistent dynamical models of six globular clusters (NGC 362, NGC 1851, NGC 2808, NGC 4372, NGC 5927 and NGC 6752). The models have been fitted to surface brightness data from \cite{Trager1995} and line-of-sight velocities from spectroscopic data (from the Gaia-ESO survey, \cite{Lardo2015}, \cite{Kacharov2014} and \cite{Lane2011}). Our work represents both an advance in theory in that the presented models are new and an advance in data analysis as we have developed a novel scheme for fitting the models to the data. We will briefly present the main achievements and conclusions of this paper.

\subsection{Action-based globular cluster models}

We found that the surface brightness profiles of our sample of globular clusters are well reproduced by a simple adjustment to the classic Plummer law. This Modified Plummer model produces a slower break in the density profile at the scale radius. Using an Eddington inversion scheme, we numerically computed the corresponding isotropic action-based distribution function and found that it was well approximated by a simple function of the actions. Flattening was introduced by scaling the vertical action by a factor $q_z$ which was found to scale linearly with the observed flattening near the scale radius $q=\tfrac{1}{2}(1+q_z)$. Rotation about the symmetry axis was introduced to the models by adding odd functions of the $z$-component of the angular momentum, which has no effect on the density profile. We implemented two rotation curves: a hyperbolic tangent model that has a flat rotation profile with angular momentum and an exponential model that limits rotation to low angular momenta so produces a rotating core.

\subsection{Data analysis}

We fitted our models to the data in two steps: first, we fixed the density profile using the surface brightness data from \cite{Trager1995}. Secondly, for each cluster we produced a grid of rotating models viewed at varying inclination angles such that the observed ellipticity matched that of each cluster. To perform inference on our parameters, we created a novel interpolation scheme for extracting arbitrary models from our grid of models. Each line-of-sight velocity distribution is expressed in terms of a q-Gaussian basis expansion and line-of-sight velocity distributions at arbitrary positions and model parameters can be reconstructed with high accuracy by linearly interpolating on the basis coefficients. We have demonstrated that this scheme works extremely well and represents an improvement over the classic Gauss-Hermite expansion as it correctly reproduces the truncation of the line-of-sight velocity distributions at the escape speed. Such an approach is applicable to analysis of any line-of-sight velocity data (from e.g. integral field spectroscopy data) and we hope that the procedure outlined in this paper will prove more generally useful.

\subsection{Results}
We have fitted our models to the globular cluster data by using MCMC to estimate the uncertainties in the derived parameters. We derived systemic velocities and mass-to-light ratios for all the clusters, which agree well with literature values using N-body models that account for effects such as mass segregation \citep{Baumgardt2017}. We also measured the velocity offsets between different datasets used in the analysis and found they agreed well with the observations of the same stars from the two datasets. 

From our model fits, we have found, in general, that there is a weak preference for the models to be viewed edge-on. We have found, from evaluating the Bayesian evidence ratio, that the hyperbolic tangent rotation model (which gives a flatter rotation curve out to infinity) produces no better a fit to all the globular clusters bar NGC 2808 than does the non-rotating model. Only in the case of NGC 2808 do we find significant evidence for rotation when fitting this flat rotation curve model. However, when fitting the exponential rotation model (which produces centralized rotation and zero rotation at infinity), we find that there is positive evidence for rotation in NGC 2808, and to a lesser extent in NGC 4372 and NGC 5927. For all other clusters, the data quality is too poor to reliably detect any rotation. We concluded our study by plotting the results of the rotation amplitude over the central dispersion against ellipticity for our sample of clusters. Our results show that there is a weak correlation between rotation signal and ellipticity, with the clusters all lying beneath the classic isotropic rotator line. This suggests that the flattening in these clusters is linked to anisotropy.

\subsection{Future directions}

The modelling presented here is the first step towards a much richer analysis. We have shown the validity of both the models and the data analysis procedure and the machinery is applicable to the many more globular clusters for which there are spectroscopic data. Our work has shown that for many globular clusters the rotation signals are weak and so more high quality data is required to robustly extract the rotation properties. Additionally, in some instances we are limited by the quality of the photometric observations used.

\cite{Watkins2015} has provided high quality proper motion measurements within globular clusters and in turn anisotropy measurements. We have used these measurements to validate our particular choice of models, but a fuller analysis would also use this data in the model fits. There is a strong link between anisotropy, rotation and flattening and so restricting the anisotropy can lead to improved measurements of the rotation. It is unlikely that the proper motion data is of high enough quality to reveal a visible rotation signal, but they may potentially provide constraints on the inclination angle.

Analysis of integral field spectroscopy data from the MUSE survey~\citep{Bacon2014} will soon yield a wealth of further kinematic measurements for globular clusters, allowing the observation of many more stars than can be observed using other spectroscopic techniques. The kinematics of NGC 6397 have already been analysed by~\cite{Kamann2016} with a sample of 12307 stars, pointing towards the possibility of a slight rotational signature in this cluster, along with slight flattening in its inner regions. The data is shown to reach an accuracy of $<1$ kms$^{-1}$, comparable to the accuracy of the measurements used in this work.

Our reliance on the position angles measured by~\cite{WhiteShawl1987} could also be eliminated in future work by using the now publicly-available photometry from the Hubble Space Telescope UV legacy survey of Galactic globular clusters \citep{Piotto2015,Soto2017}. By engaging with the photometric data in two dimensions rather than fitting the density profile along one radial trajectory, the density isophotes of our flattened models could be used to produce a separate prediction for the position angle of each cluster.

Finally, we opened our paper with a discussion of the potential importance of dynamical models in distinguishing between models of globular cluster formation and evolution. In particular, we mentioned that action-based models are ideally suited to dealing with multiple populations, as we are free to sum an arbitrary number of components without changing our algorithms. A necessary next step is to separate the globular cluster into subpopulations and simultaneously fit a model for each component. Many globular clusters exhibit Na-O anti-correlation signatures~\citep{Carretta2010}, such that an interesting first step would be to separate the data from our clusters into $[\mathrm{Na}/\mathrm{O}]$ rich and poor populations and fit a sum of two of the presented models to detect any correlations between chemistry and dynamics \citep{Bellazzini2012}. \cite{Cordero2017} has already demonstrated the feasibility of such a study using M 13. Such developments will bring globular clusters in line with dwarf spheroidal galaxies or components of the Milky Way, for which chemo-dynamical modelling is now routine.

\section*{Acknowledgments}
Based on data products from observations made with ESO Telescopes at the La Silla Paranal Observatory under programme ID 188.B-3002. These data products have been processed by the Cambridge Astronomy Survey Unit (CASU) at the Institute of Astronomy, University of Cambridge, and by the FLAMES/UVES reduction team at INAF/Osservatorio Astrofisico di Arcetri. These data have been obtained from the Gaia-ESO Survey Data Archive, prepared and hosted by the Wide Field Astronomy Unit, Institute for Astronomy, University of Edinburgh, which is funded by the UK Science and Technology Facilities Council. This work was partly supported by the European Union FP7 programme through ERC grant number 320360 and by the Leverhulme Trust through grant RPG-2012-541. We acknowledge the support from INAF and Ministero dell' Istruzione, dell' Universit\`a' e della Ricerca (MIUR) in the form of the grant "Premiale VLT 2012". The results presented here benefit from discussions held during the Gaia-ESO workshops and conferences supported by the ESF (European Science Foundation) through the GREAT Research Network Programme.

\bibliographystyle{mnras}
\bibliography{bibliography}

\appendix
\section{Surface density profile of modified Plummer law}\label{Appendix::SurfaceDensity}
In equation~\eqref{Eqn::NewPlummer}, we introduced a density profile that bears a resemblance to the classic Plummer profile in that its asymptotic limits are $\rho\sim r^0$ and $\rho\sim r^{-5}$, but it has a slower turn-over between these regimes. The corresponding 3D mass profile $m(r)$ is given by
\begin{equation}
m(r)/M = 1-\frac{1+4r+6r^2}{(1+r)^4}.
\end{equation}
Solving $m(r)/M=\tfrac{1}{2}$ gives a solution for the 3D half-light radius $r_h=1.5925r_s$. This profile also has an analytic surface density profile given by
\begin{equation}
\Sigma(R) = \frac{M}{4\pi r_s^2}\frac{(16s^4+83s^2+6)-15s^2(3s^2+4)X(s)}{(s^2-1)^4},
\end{equation}
where $s=R/r_s$ and $X(s)$ is the function introduced by \cite{Hernquist1990} given by
\begin{equation}
X(s)=
\begin{cases}
\mathrm{arcsech}(s)/\sqrt{1-s^2}& \text{ if } 0\leq s\leq1  \\
\mathrm{arcsec}(s)/\sqrt{s^2-1}& \text{ if } s\geq1,
\end{cases}
\end{equation}
which is continuous through $s=1$. Similarly, the mass contained within a cylinder of radius $R$, $M'(R)$, is given by
\begin{equation}
M'(R)/M = 1+\tfrac{1}{2}\frac{(2-9s^2-8s^4)+15s^4X(s)}{(s^2-1)^3}.
\end{equation}
The numerical solution for $R$ of $M'(R)/M=\tfrac{1}{2}$ results in $R=R_h=1.2038r_s$. Comparison to the Plummer profile where $R_h=r_s$ and $r_h=1.3048r_s$ we see that as expected from a slower turn-over, the half-light radii for our model lie further out than in the corresponding Plummer case.

\section{q-Gaussian polynomial basis expansion}\label{Appendix::qGauss}
Line profiles of galaxies are often simplified through a representation in terms of Gauss-Hermite polynomials $H_i(y)$ \citep{vanderMarel1993, Magorrian1994}. 
Such an expansion is valid for distributions that extend to $\pm\infty$ but is inappropriate for distributions that truncate at some fixed value. 
For instance, when representing the l.o.s. velocity profiles of our models, we require the weight to be identically zero beyond the escape velocity $V_\mathrm{esc}$. In this appendix we will give a set of orthogonal polynomials using the q-Gaussian as a weight function which are appropriate for this problem.

The unit q-Gaussian weight function is defined as
\begin{equation}
\mathcal{W}(y;q)=C_q (1-(1-q)y^2)^{1/(1-q)},
\end{equation}
where $C_q$ is a normalization constant given by
\begin{equation}
C_q = \sqrt{\frac{1-q}{\pi}}\frac{\GammaAB{5-3q}{2(1-q)}}{\GammaAB{2-q}{1-q}}.
\end{equation}
For $q<1$, $\mathcal{W}=0$ at $y=\pm y_0=\pm(1-q)^{-1/2}$. Additionally, we define the constant
\begin{equation}
\mathcal{N}_q = \int_{-y_0}^{y_0}\dd y\,\mathcal{W}^2(y;q) = C_q^2\sqrt{\frac{\pi}{1-q}}\frac{\GammaAB{3-q}{1-q}}{\GammaAB{7-3q}{2(1-q)}}.
\end{equation}

We use Gram-Schmidt orthogonalization to construct a set of orthonormal polynomials $\mathcal{Q}_i(y)$ on the interval $(-y_0,y_0)$. The polynomials are normalized such that
\begin{equation}
\int_{-y_0}^{y_0}\dd y\,\mathcal{W}^2(y;q)\mathcal{Q}^2_i(y;q) = \mathcal{N}_q.
\end{equation}

The first five polynomials are given by
\begin{equation}
\begin{split}
\mathcal{Q}_0(y;q)=&1,\\
\mathcal{Q}_1(y;q)=&x\sqrt{7-3q},\\
\mathcal{Q}_2(y;q)=&\frac{1}{2}\sqrt{\frac{9-5q}{3-q}}[(7-3q)x^2-1],\\
\mathcal{Q}_3(y;q)=&\frac{1}{2}\frac{9-5q}{\sqrt{9-3q}}\sqrt{(7-3q)(11-7q)}[x^3-\frac{3}{9-5q}x],\\
\mathcal{Q}_4(y;q)=&\sqrt{\frac{(7-3q)(13-9q)}{192(3-q)(2-q)}}\times\\&[(9-5q)(11-7q)x^4-6(9-5q)x^2+3].
\end{split}
\end{equation}

Given some function $f(y)$, the coefficients of the q-Gaussian polynomial expansion are given by
\begin{equation}
\mathfrak{q}_i = \mathcal{N}_q^{-1}\int_{-y_0}^{y_0}\dd y\,f(y)\mathcal{W}(y;q)\mathcal{Q}_i(y;q),
\end{equation}
and the representation of $f$ can be reconstructed as
\begin{equation}
\hat{f}(y)=\mathcal{W}(y;q)\sum_i \mathfrak{q}_i \mathcal{Q}_i(y;q).
\end{equation}

\subsection{Choice of q}
For the best representation of a general profile $F(x)$, we must first shift and scale the ordinate of the distribution such that the peak at $x=x_c$ coincides with $y=0$ and the width is approximately that of the unit q-Gaussian. We opt to match the standard deviation of $F$ to the standard deviation of the q-Gaussian given by $\sigma_q = 1/(5-3q)$. If $F(\pm x_e)=0$, we define $\sigma$ as the standard deviation of $F$ and set $x_0=\mathrm{max}(x_e-x_c,x_c-x_e)$. Then an appropriate scaling is given by
\begin{equation}
\beta = \sqrt{\frac{1}{2\sigma^2}-\frac{3}{2 x_0^2}},
\end{equation}
and
\begin{equation}
f(y) = F(y/\beta+x_c).
\end{equation}
Additionally, we choose $q$ as
\begin{equation}
q = 1 - \frac{1}{(\beta x_0)^2}.
\end{equation}

\subsection{Checks of the q-Gaussian interpolation}
\begin{figure*}
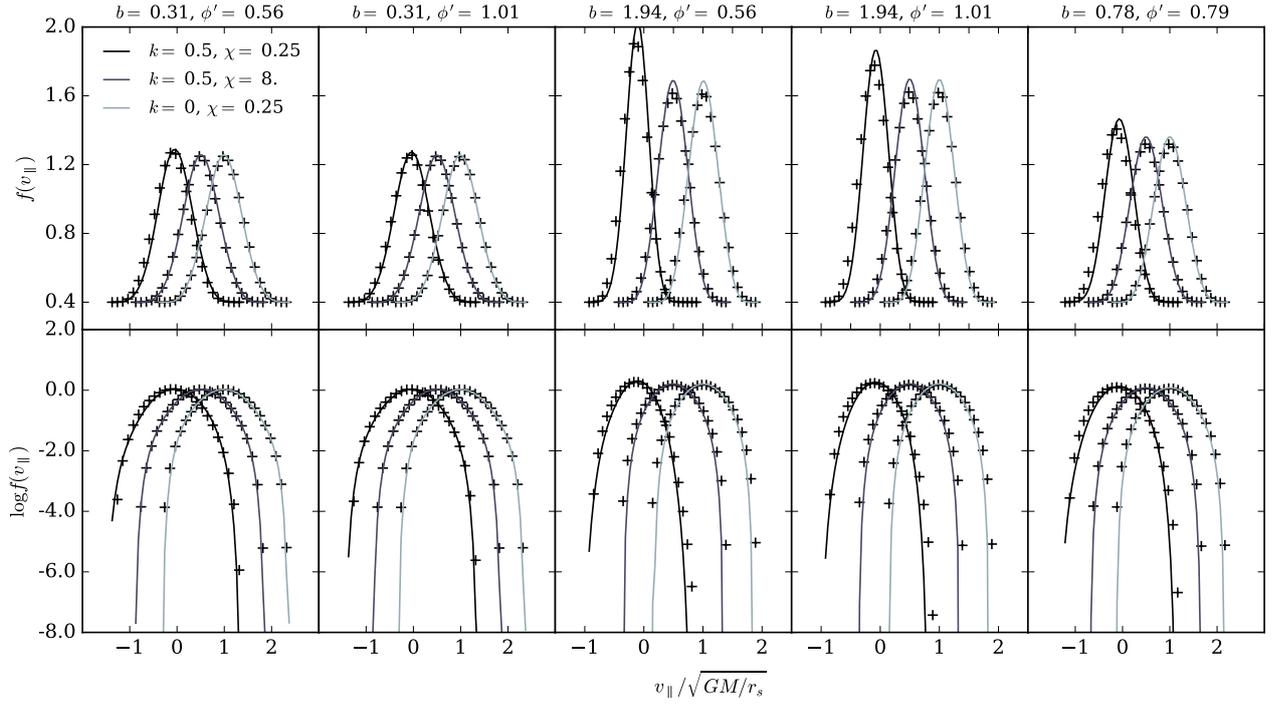

$$\includegraphics[width=0.95\textwidth]{{{figs/FigB1}}}$$
\caption{Comparison between q-Gaussian Hermite fits interpolated at five points between the $(b,\phi^{\prime})$ grid points of Sec.~\ref{Sec::Model} and the l.o.s. velocity distributions obtained by computing the self-consistent model at these points in parameter space, for the most spherical model $(q_z,\theta)=(0.98,1.57)$. We see that the q-Gaussian Hermite distributions interpolate well over the position co-ordinates.}
\label{Fig::interpolation_posn1}
\end{figure*}

In Section~\ref{Section::ModelRepresentation} we presented some checks of the accuracy of the q-Gaussian expansion scheme by comparing the reconstructed line-of-sight velocity distributions at several on-sky positions and also as a function of the rotation amplitude $k$. In this appendix, we present a much fuller range of checks of the accuracy of our procedure. In Fig.~\ref{Fig::interpolation_posn1}, we show the actual distributions alongside the reconstructed distributions for five different on-sky positions for three different near-spherical models (two maximally rotating exponential models with $\chi=0.25\sqrt{GMr_s}$ and $\chi=8\sqrt{GMr_s}$ and a non-rotating model) that lie on the interpolation grid points. This procedure tests the quality of the on-sky interpolation. We see that the q-Gaussians capture the shapes of the distributions over many orders of magnitude and possibly the only criticism is the small failure near the peaks of the distributions. In Fig.~\ref{Fig::interpolation_posn2}, we show the same results for the most flattened model considered $q_z=0.6$ and again the results are very satisfactory and as good as the near-spherical case. 

In Figures~\ref{Fig::interpolation_model_qz095} and~\ref{Fig::interpolation_model_qz065} we show the comparison of the actual and the reconstructed distributions for the most spherical and most flattened models respectively. We show the results evaluated at two on-sky positions for a range of different rotation curve parameters. For the most flattened case we see the inability of the q-Gaussian expansion to correctly capture the height of the peak for the most extreme on-sky position. However, we note that these tests are at extreme values and so not representative of the whole. Indeed, the figures demonstrate the high quality with which we can reproduce the distribution anywhere and for any model.

\begin{figure*}
$$\includegraphics[width=0.95\textwidth]{{{figs/FigB2}}}$$
\caption{Comparison between q-Gaussian Hermite fits interpolated at five points between the $(b,\phi^{\prime})$ grid points of Sec.~\ref{Sec::Model} and the l.o.s. velocity distributions obtained by computing the self-consistent model at these points in parameter space, for the most flattened model $(q_z,\theta)=(0.6,0.77)$. We see that the q-Gaussian Hermite distributions interpolate well over the position co-ordinates.}
\label{Fig::interpolation_posn2}
\end{figure*}

\begin{figure*}
$$\includegraphics[width=0.91\textwidth]{{{figs/FigB3}}}$$
\caption{Comparison between q-Gaussian Hermite fits interpolated at $(q_z,\theta)=(0.95,0.63)$ and at four values of $(q_z,k,\chi)$, and the l.o.s. velocity distributions obtained by computing the self-consistent model at these points in parameter space. The two different values of $b=1.34,12$ tested are the grid-points of maximum rotation and of maximum distance from the cluster centre, respectively. We hold $\phi^{\prime}=0.1$ such that the effects of rotation can be most clearly observed. We see that the q-Gaussian Hermite distributions interpolate well over the physical model parameters, particularly in logarithmic space.}
\label{Fig::interpolation_model_qz095}
\end{figure*}

\begin{figure*}
$$\includegraphics[width=0.91\textwidth]{{{figs/FigB4}}}$$
\caption{Comparison between q-Gaussian Hermite fits interpolated at $(q_z,\theta)=(0.65,0.22)$ and at four values of $(q_z,k,\chi)$, and the l.o.s. velocity distributions obtained by computing the self-consistent model at these points in parameter space. The two different values of $b=1.34,12$ tested are the grid-points of maximum rotation and of maximum distance from the cluster centre, respectively. We hold $\phi^{\prime}=0.1$ such that the effects of rotation can be most clearly observed. We see that the q-Gaussian Hermite distributions interpolate well over the physical model parameters, particularly in logarithmic space.}
\label{Fig::interpolation_model_qz065}
\end{figure*}

\bsp
\label{lastpage}

\end{document}

%% file: Table1.dat
\begin{minipage}{\textwidth}
\centering
\begin{tabular}{lccccccccccc}
Name&$N$&$N_{\mathrm{GES}}$&$N_{\mathrm{Lardo}}$&$N_\mathrm{Other}$&$R_{\text{5th}}/r_h$&$R_{\text{95th}}/r_h$&$d$&$\epsilon_\mathrm{W\&S}$&P.A.$_{\text{W\&S}}$&P.A.$_{\text{rot}}$&$A_0$\\
&&&&&&&$[\mathrm{kpc}]$&&$[\mathrm{deg}]$&$[\mathrm{deg}]$&$[\mathrm{km\,s}^{-1}]$\\
&(1)&(2)&(3)&(4)&(5)&(6)&(7)&(8)&(9)&(10)&(11)\\
\hline
NGC 362 & 122 & 122 & 0 & 0 & 1.14 & 8.02 & 8.6 & 0.01 (2.15) & 25 $\pm$ 4 & 131 $\pm$ 12 & $1.3\pm0.5$ \\
NGC 1851 & 221 & 64 & 157 & 0 & 1.64 & 14.21 & 12.1 & 0.05 (2.42) & 86 $\pm$ 1 & 116 $\pm$ 8 & $1.0\pm0.2$ \\
NGC 2808 & 117 & 42 & 75 & 0 & 0.96 & 10.65 & 9.6 & 0.12 (2.83) & 31 $\pm$ 0 & 3 $\pm$ 5 & $4.0\pm0.6$ \\
NGC 4372\footnote{For NGC 4372, we use the ellipticity ($\epsilon=0.08$) and position angle ($48\,\mathrm{deg}$) measured by \protect\cite{Kacharov2014}.} & 196 & 45 & 111 & 40 & 0.30 & 2.02 & 5.8 & 0.15 (0.64) & 38 $\pm$ 6 & 48 $\pm$ 19 & $0.6\pm0.3$\\
NGC 5927 & 87 & 33 & 54 & 0 & 0.70 & 6.92 & 7.7 & 0.04 (1.15) & 63 $\pm$ 2 & 132 $\pm$ 9 & $1.9\pm0.4$ \\
NGC 6752 & 884 & 209 & 503 & 172 & 0.45 & 5.85 & 4.0 & 0.04 (1.24) & 147 $\pm$ 7 & 125 $\pm$ 17 & $0.3\pm0.1$ \\
\hline
\end{tabular}
\end{minipage}

%% file: Table2.dat
\begin{tabular}{llcccccccc}
Model&Name&$\cos\theta$&$k$&$\log_2(\frac{\chi}{\sqrt{GMr_s}})$&$v_\mathrm{sys}$&$\Upsilon$&$\Delta_2$&$\Delta_3$&$2\ln K$\\
&&&&&$[\mathrm{km\,s}^{-1}]$&$[\Upsilon_\odot]$&$[\mathrm{km\,s}^{-1}]$&$[\mathrm{km\,s}^{-1}]$&\\
\hline\\
norot&362&$0.45^{+0.29}_{-0.29}$&-&-&$222.75^{+0.46}_{-0.41}$&$1.97^{+0.20}_{-0.17}$&-&-&-\\
&1851&$0.39^{+0.33}_{-0.28}$&-&-&$319.81^{+0.66}_{-0.53}$&$1.98^{+0.15}_{-0.11}$&$0.04^{+0.64}_{-0.65}$&-&-\\
&2808&$0.18^{+0.16}_{-0.13}$&-&-&$101.61^{+0.72}_{-1.07}$&$2.45^{+0.16}_{-0.13}$&$2.07^{+1.09}_{-0.94}$&-&-\\
&4372&$0.38^{+0.28}_{-0.27}$&-&-&$74.82^{+0.61}_{-0.39}$&$2.05^{+0.11}_{-0.12}$&$0.21^{+0.82}_{-0.64}$&$1.57^{+0.75}_{-0.91}$&-\\
&5927&$0.22^{+0.19}_{-0.16}$&-&-&$-103.27^{+0.34}_{-0.34}$&$2.94^{+0.17}_{-0.18}$&$-1.01^{+0.50}_{-0.44}$&-&-\\
&6752&$0.21^{+0.24}_{-0.14}$&-&-&$-26.23^{+0.20}_{-0.27}$&$2.24^{+0.07}_{-0.08}$&$-0.67^{+0.31}_{-0.37}$&$0.12^{+0.41}_{-0.41}$&-\\
\hline\\
tanh&362&$0.47^{+0.30}_{-0.35}$&$0.16^{+0.18}_{-0.29}$&$1.22^{+1.06}_{-2.50}$&$222.79^{+0.54}_{-0.48}$&$2.02^{+0.22}_{-0.18}$&-&-&$0.00\pm0.18$\\
&1851&$0.56^{+0.22}_{-0.40}$&$0.02^{+0.14}_{-0.23}$&$0.80^{+1.79}_{-2.74}$&$319.84^{+0.50}_{-0.50}$&$1.96^{+0.16}_{-0.18}$&$-0.02^{+0.54}_{-0.75}$&-&$-0.17\pm0.22$\\
&2808&$0.23^{+0.16}_{-0.13}$&$0.30^{+0.09}_{-0.10}$&$-0.91^{+2.29}_{-0.84}$&$101.59^{+1.13}_{-0.89}$&$2.46^{+0.15}_{-0.15}$&$2.05^{+1.05}_{-1.17}$&-&$4.06\pm0.28$\\
&4372&$0.41^{+0.23}_{-0.22}$&$0.34^{+0.11}_{-0.16}$&$-0.39^{+1.62}_{-1.10}$&$74.90^{+0.49}_{-0.33}$&$2.08^{+0.16}_{-0.20}$&$0.15^{+0.43}_{-0.55}$&$2.02^{+0.56}_{-0.76}$&$1.73\pm0.28$\\
&5927&$0.26^{+0.19}_{-0.17}$&$0.02^{+0.22}_{-0.23}$&$1.26^{+1.13}_{-2.22}$&$-103.29^{+0.45}_{-0.34}$&$2.94^{+0.14}_{-0.10}$&$-1.13^{+0.57}_{-0.50}$&-&$-0.52\pm0.30$\\
&6752&$0.27^{+0.18}_{-0.19}$&$-0.14^{+0.18}_{-0.11}$&$0.32^{+1.80}_{-1.44}$&$-26.25^{+0.29}_{-0.29}$&$2.24^{+0.11}_{-0.07}$&$-0.68^{+0.42}_{-0.33}$&$0.03^{+0.41}_{-0.37}$&$-0.31\pm0.33$\\
\hline\\
exp&362&$0.40^{+0.37}_{-0.24}$&$0.14^{+0.22}_{-0.28}$&$0.70^{+1.89}_{-1.79}$&$222.87^{+0.47}_{-0.55}$&$2.03^{+0.16}_{-0.15}$&-&-&$-0.07\pm0.18$\\
&1851&$0.51^{+0.22}_{-0.41}$&$-0.07^{+0.29}_{-0.17}$&$0.25^{+2.30}_{-1.81}$&$319.79^{+0.57}_{-0.72}$&$1.98^{+0.14}_{-0.15}$&$0.07^{+0.67}_{-0.71}$&-&$-0.14\pm0.22$\\
&2808&$0.16^{+0.13}_{-0.11}$&$0.37^{+0.08}_{-0.10}$&$-0.43^{+1.84}_{-1.01}$&$101.65^{+0.84}_{-0.63}$&$2.42^{+0.11}_{-0.13}$&$1.85^{+0.97}_{-1.00}$&-&$7.19\pm0.28$\\
&4372&$0.47^{+0.24}_{-0.32}$&$0.23^{+0.15}_{-0.15}$&$-0.15^{+2.02}_{-2.09}$&$74.86^{+0.56}_{-0.76}$&$2.05^{+0.12}_{-0.15}$&$0.30^{+0.72}_{-0.71}$&$1.78^{+1.01}_{-0.80}$&$2.37\pm0.28$\\
&5927&$0.13^{+0.23}_{-0.10}$&$0.17^{+0.19}_{-0.35}$&$-1.38^{+1.59}_{-0.46}$&$-103.26^{+0.46}_{-0.41}$&$3.00^{+0.12}_{-0.18}$&$-0.89^{+0.50}_{-0.58}$&-&$1.44\pm0.30$\\
&6752&$0.20^{+0.24}_{-0.13}$&$-0.14^{+0.07}_{-0.07}$&$-0.10^{+1.74}_{-0.96}$&$-26.19^{+0.32}_{-0.30}$&$2.22^{+0.11}_{-0.08}$&$-0.78^{+0.37}_{-0.36}$&$0.08^{+0.54}_{-0.31}$&$0.01\pm0.34$\\
\hline\\
\end{tabular}

%% file: Table3.dat
\begin{minipage}{\columnwidth}
\begin{tabular}{lcccc}
Name&$\Upsilon_{\text{MVdM}}/\Upsilon_{\odot}$&$\Upsilon_{\text{W,K}}/\Upsilon_{\odot}$&$\Upsilon_{\text{B}}/\Upsilon_{\odot}$&$\Upsilon_{\text{tw}}/\Upsilon_{\odot}$ \\
& (1) & (2) & (3) & (4) \\
\hline \\
362 & - & - & $1.73\pm0.26$ & $1.97^{+0.20}_{-0.17}$ \\
1851 & $1.98$ & $1.51 \pm 0.03$ & $2.40\pm0.2$ & $1.98^{+0.15}_{-0.11}$ \\
2808 & $2.02$ & $1.56 \pm 0.02$ & $1.96\pm0.16$ & $2.42^{+0.11}_{-0.13}$ \\
4372 & - & $1.7 \pm 0.4$\footnote{For NGC 4372, we quote the value measured by \protect\cite{Kacharov2014}.} & $1.67\pm0.19$ & $2.05^{+0.11}_{-0.12}$ \\
5927 & $2.93$ & $1.48^{+0.15}_{-0.17}$  & $2.19\pm0.42$ & $2.94^{+0.17}_{-0.18}$ \\
6752 & $1.88$ & $2.14^{+0.05}_{-0.06}$ & $2.60\pm0.41$&$2.24^{+0.07}_{-0.08}$ \\
\hline \\
\end{tabular}
\end{minipage}

%% file: Table4.dat
\begin{tabular}{lcccc}
Name&$\Delta_2$&$\Delta_{2,\mathrm{dupl}}$&$\Delta_3$&$\Delta_{3,\mathrm{dupl}}$\\
\hline\\
1851&$0.04^{+0.64}_{-0.65}$&$-0.22^{+0.33}_{-0.35}$&-&-\\2808&$2.07^{+1.09}_{-0.94}$&$-0.04^{+0.16}_{-0.32}$&-&-\\4372&$0.21^{+0.82}_{-0.64}$&$-0.04^{+0.21}_{-0.23}$&$1.57^{+0.75}_{-0.91}$&$0.49^{+0.47}_{-0.49}$\\5927&$-1.01^{+0.50}_{-0.44}$&$-0.09^{+0.17}_{-0.13}$&-&-\\6752&$-0.67^{+0.31}_{-0.37}$&$-0.73^{+0.59}_{--0.64}$&$0.12^{+0.41}_{-0.41}$&$0.42^{+0.72}_{-0.90}$\\\hline\\
\end{tabular}